\documentclass[namedreferences]{solarphysics}

\usepackage{adjustbox}
\usepackage{amsfonts}

\newcommand{\arcsec}{\hbox{$^{\hbox{\rlap{\hbox{\lower4pt\hbox{$\,\prime\prime$}}
          }\hbox{$\frown$}}}$}}
\usepackage[hyperref,optionalrh,solaromanenum]{spr-sola-addons} % For Solar Physics 
\usepackage{graphicx}                    % For eps figures, newer & more powerfull
\usepackage{color}                       % For color text: \color command
\usepackage{natbib}
\def\UrlFont{\sf}                        % define the fonts for the URLs
\newcommand{\aap}{    {\it Astron. Astrophys.}}

\newcommand{\aapr}{   {\it Astron. Astrophys. Rev.}}

\newcommand{\apj}{    {\it Astrophys. J.}}
\newcommand{\apjs}{    {\it Astrophys. J. Suppl.}}
\newcommand{\apjl}{   {\it Astrophys. J. Lett.}}

\newcommand{\jastp}{  {\it J. Atmos. Solar-Terr. Phys.}} 
\newcommand{\jgr}{    {\it J. Geophys. Res.}}

\newcommand{\solphys}{{\it Solar Phys.}}

\chardef\us=`\_

\begin{document}

\begin{article}

\begin{opening}

\title{A Comparative Study of the Eruptive and Non-Eruptive Flares Produced by the Largest Active Region of Solar Cycle 24  }

%%%%%%%%%%%%%%%%%%%%%%%%%%%%%%%%%%%%%%%%%%%%%%%%%%%
%Authors Names
%
\author[addressref={1},corref,email={ranadeep@prl.res.in}]{\inits{Ranadeep Sarkar}\fnm{Ranadeep Sarkar}\lnm{}}
\author[addressref={1}]{\inits{Nandita Srivastava}\fnm{Nandita Srivastava}\lnm{}}
%%%%%%%%%%%%%%%%%%%%%%%%%%%%%%%%%%%%%%%%%%%%%%%%%%%
%% Runningheads
%
\runningauthor{R.Sarkar, N.Srivastava}
\runningtitle{A Comparative Study of Confined and Eruptive Flares Produced by AR 12192}

%%%%%%%%%%%%%%%%%%%%%%%%%%%%%%%%%%%%%%%%%%%%%%%%%%%
%% Affilations 
%% id shold be the same with \author addressref value.
\address[id={1}]{Udaipur Solar Observatory, Physical Research Laboratory, Box 198, Badi Road, Udaipur 313001, India}

%%%%%%%%%%%%%%%%%%%%%%%%%%%%%%%%%%%%%%%%%%%%%%%%%%%
%%% Abstract 
\begin{abstract}
We investigate the morphological and magnetic characteristics of  solar active region (AR) NOAA 12192. AR 12192 was the largest region of Solar Cycle 24; it underwent noticeable growth and produced 6 X-class flares, 22 M-class flares, and 53 C-class flares in the course of its disc passage. However, the most peculiar fact of this AR is that it was associated with only one CME in spite of producing several X-class flares. In this work, we carry out a comparative study between the eruptive and non-eruptive flares produced by AR 12192. We find that the magnitude of abrupt and permanent changes in the horizontal magnetic field and Lorentz force are significantly smaller in the case of the confined flares compared to the eruptive one. We present the areal evolution of AR 12192 during its disc passage. We find the flare-related morphological changes to be weaker during the confined flares, whereas the eruptive flare exhibits a rapid and permanent disappearance of penumbral area away from the magnetic neutral line after the flare. Furthermore, from the extrapolated nonlinear force-free magnetic field, we examine the overlying coronal magnetic environment over the eruptive and non-eruptive zones of the AR. We find that the critical decay index for the onset of torus instability was achieved at a lower height over the eruptive flaring region, than for the non-eruptive core area. These results suggest that the decay rate of the gradient of overlying magnetic field strength may play a decisive role to determine the CME productivity of the AR. In addition, the magnitude of changes in the flare-related magnetic characteristics are found to be well correlated with the nature of solar eruptions.

\end{abstract}

%%%%%%%%%%%%%%%%%%%%%%%%%%%%%%%%%%%%%%%%%%%%%%%%%%%
%% Keywords
%

\end{opening}
%-------------------------------------------------

%%%%%%%%%%%%%%%%%%%%%%%%%%%%%%%%%%%%%%%%%%%%%%%%%%%
%% Sections

\section{Introduction}
Solar flares and coronal mass ejections (CMEs) are believed to be the most violent and energetic phenomena that occur in the solar atmosphere. Together they can release highly energetic particle radiation and gigantic clouds of ionized gas that may have severe impacts on the human high-tech activities in outer space \citep{Gosling,Siscoe,Daglis}. Several studies have been made to understand the physical processes behind the origin of CMEs and solar flares \citep{Kahler,Gosling}. Nevertheless,  the initiation mechanism of CMEs and their association with flares still remain as one of the most elusive topics in solar physics \citep{Schrijver}. 

 Earlier observations reveal that flares and CMEs can be regarded as two different manifestations of the same energy release process \citep{harri,zhang,son1}. \citet{zhang} illustrated that the fast acceleration phase of CMEs in the inner corona is temporally correlated with the rise time of the associated soft X-ray flares, suggesting both of the phenomena to be connected through the same physical process, possibly via magnetic reconnection \citep{LinForbes,PriestForbes}. Despite the intrinsic physical relationship between flares and CMEs, observations reveal that not all flares are associated with CMEs \citep{Andrews,Yashiro2005}.  

Several attempts have been made to explore the possible circumstances under which solar flares may lead to failed eruptions. From a comparative study between four eruptive and four non-eruptive X-class flares produced from different active regions (AR), \citet{WangZang} found that the confined flares tend to occur close to the magnetic center of the AR whereas the eruptive ones generally occur away from the magnetic center. \citet{TorokKliem}, \citet{KliemTorok}, \cite{FanGibson}, and \cite{OlmedoZhang} investigated the conditions leading to instability of the flux rope structures in the context of the torus instability, which is one of the ideal MHD instabilities \citep{Priest2014}. They found that the gradient of the overlying magnetic field strength exhibits a critical limit that may determine whether the flux rope will result in eruption or not. Furthermore, \citet{Liu2008} carried out a comparative study among the ten events including four failed eruptions (FE), four eruptions due to torus instability (TI) and two eruptions due to kink instability (KI) from different ARs.  His results led him to speculate that the gradient of overlying magnetic field strength decays faster in TI and KI driven events in comparison to the FE events.

The above results depict the role of background magnetic-field strength in determining the confinement behavior of the solar eruptions. To further investigate the distinct properties of confined and eruptive events, we  perform a comparative study of the eruptive and non-eruptive flares produced by AR 12192, which was the largest active region of Solar Cycle 24. However, the most peculiar aspect of this AR was that it produced six X-class flares during its disc passage but none of them were associated with CMEs. However, from a study of flare-CME association rate it was found that 75 $\%$ of GOES flare $\geq$ X1.0 class are CME productive, and for the flare class $\geq$ X2.5 the CME association rate is more than 90 $\%$ \citep{Yashiro2006}. Thus the flare-rich but CME-poor AR 12192 drew considerable attention from the solar community \citep{Sun2015,Chen2015, Thalmann2015,Lijuan2016,Jiang}. 
\begin{figure} [!t]
\centering
  \includegraphics[width=\textwidth]{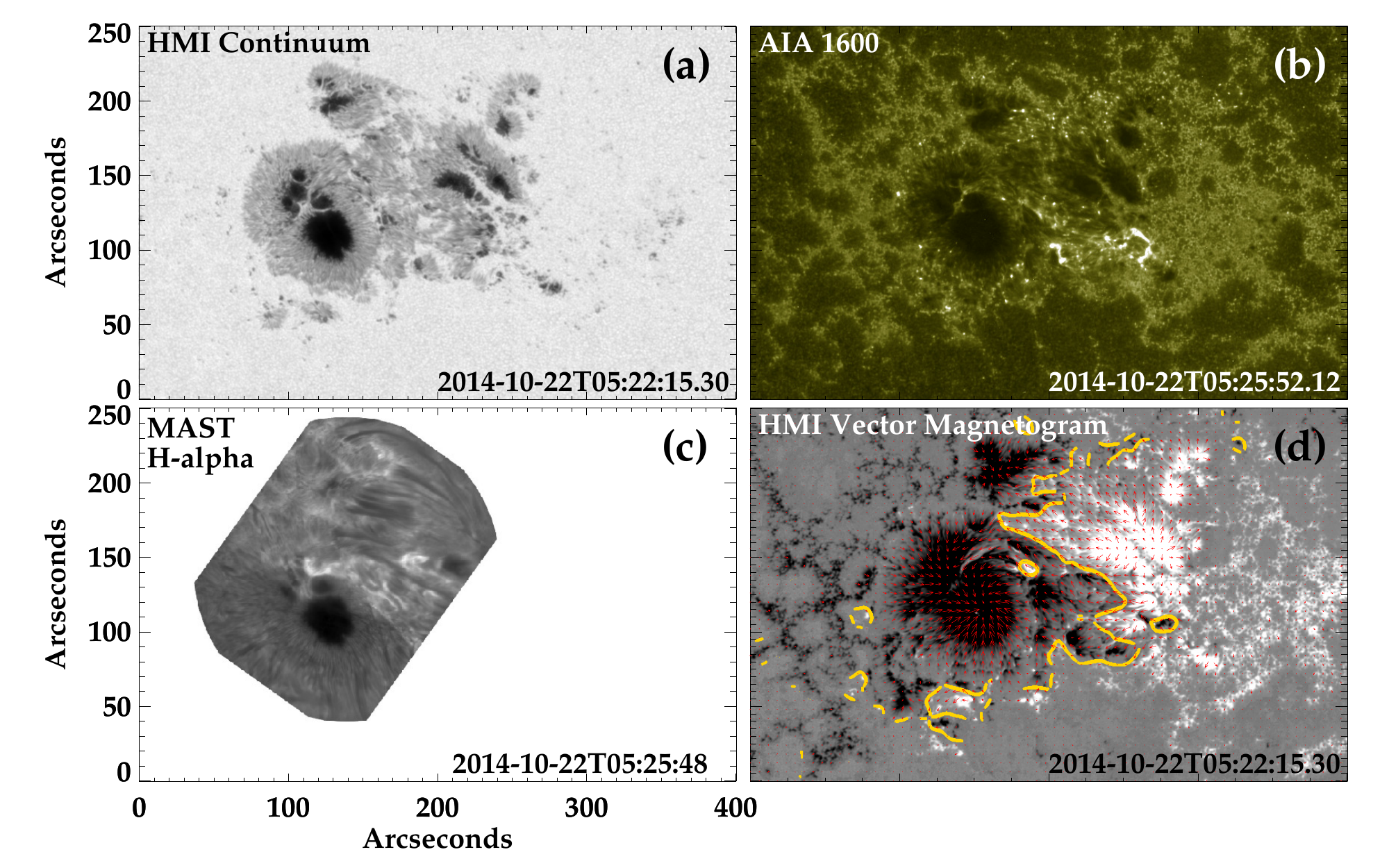}
  \caption{\textit{Panel a}: HMI continuum intensity-map of AR 12192.\textit{ Panel b}: AR 12192 in AIA 1600 \AA $ $ image.\textit{ Panel c}: high resolution H$\alpha$ image of AR 12192 taken from Multi Application Solar Telescope (MAST) at the Udaipur Solar Observatory. Due to the large size of AR 12192, the whole AR could not be captured within the field of view (3 arc-minutes) of MAST.\textit{ Panel d}: HMI vector magnetogram of the AR. The radial component [$B_{\rm r}$] of magnetic field is shown in gray scale and the horizontal component [$B_{\rm h}$] of that by red arrows, with saturation values $\pm$1000 G. The yellow lines in the panel illustrate the polarity inversion line.}
\end{figure} 

AR 12192 crossed the visible solar disc from 17 to 30 October 2014 and produced 6 X-class flares, 22 M-class flares, and 53 C-class flares. Notably, all of the energetic confined flares produced by AR 12192 occurred in the core region of that AR. On 24 October, this AR produced one X3.1-class flare, which became a record-setting event in flare energy associated with a confined eruption (\citealp{RHESSI}). As the location (S16W21) of the AR was very close to the solar disc center during that high energetic X3.1-class flare, an Earthward-directed halo CME event was expected. However, no CME was observed. Throughout the whole disc passage only one CME was produced by this AR, during an M4.0-class flare that occurred away from the core region on 24 October. \citet{Li2015} reported that this eruptive flare occurred close to the open-field-line region in which the large coronal loops, appearing as extreme ultraviolet (EUV) structures, fan out rapidly. They suggested that the interaction between the flare material and the neighboring open field lines may have caused the M4.0-class flare to become eruptive.

 The production of several non-eruptive highly energetic X-class flares, as well as the CME-associated M4.0-class flare, made the AR 12192 a unique target for a comparative study of the eruptive and non-eruptive flares originating from the same AR. In this work, we particularly aim to study the flare related changes in morphological and magnetic characteristics associated with both the confined and eruptive flares produced by AR 12192.  In order to understand the changes in photospheric magnetic field related to flares, several studies have been made in the past \citep{SudolHarvey,Petrie2010,Wang,Bkumar}. \citet{WangLiu} found an increase of transverse field at the polarity inversion line (PIL) for 11 X-class flares. \citet{Petrie} analyzed six major flares of four active regions, NOAA 11158, 11166, 11283, and 11429, and he found an increase in field strength in each case at the time of the flare, particularly in its transverse component close to the polarity inversion line (PIL). This increase in the transverse component of photospheric magnetic field was accompanied by a large, abrupt, downward vertical Lorentz-force change. \citet{Bkumar} suggested that the flare-associated Lorentz-force impulse could power the photospheric and chromospheric oscillations. A detailed physical explanation of these was first given by \citet{FletcherHudson}. However, the flare-related photospheric magnetic field and Lorentz-force changes in the case of confined flares are poorly studied. In that way, AR 12192 gave a unique opportunity to verify and extend the above results for the several confined flares produced by it. In addition we examine the pre-flare overlying coronal magnetic environment for both the eruptive and non-eruptive zone of the AR in the context of torus instability. In earlier studies, using a potential-field-extrapolation model, \citet{Thalmann2015} found a similar strong constraint on the overlying magnetic field for both the eruptive and non-eruptive flares. However, in this work we have extrapolated the photospheric magnetic field using non linear force free field (NLFFF) \citep{Aschwanden} extrapolation, which is believed to be more realistic than the potential-field-extrapolation technique \citep{Schrijver}. In Section 2, we present the data analysis and methodology. The distinct properties of the confined and eruptive flares are given in Section 3. Finally, we summarize our results in Section 4.

\section{Data Analysis}%\label{s:?}

The evolution of AR 12192 in the course of its disc passage was well recorded by the observations from the Atmospheric Imaging Assembly (AIA; \citealt{Lemen}) and the Helioseismic and Magnetic Imager (HMI; \citealt{Schou}) onboard Solar Dynamics Observatory (SDO; \citealt{Pesnell}).

To study the photospheric magnetic-field evolution of AR 12192, we have used the HMI vector magnetogram series from the version of Space weather HMI Active Region Patches (SHARP; \citealt{Turmon}) having spatial resolution of 0.5$''$ and 12 minute temporal cadence. The Stokes parameters I, Q, U, and V were derived from the filtergrams of six polarization states at six wavelengths centered on the Fe I 617.3 nm spectral line and were inverted using the Very Fast Inversion of the Stokes Algorithm code \citep{Borrero} to obtain the vector magnetic-field components in the photosphere. The remaining 180$^{\circ} $ ambiguity in the azimuthal field component was resolved using the minimum energy method \citep{Metcalf,Leka}. A coordinate transformation for remapping the vector fields onto the Lambert cylindrical equal area projection was carried out, and finally the vector fields were transformed into heliocentric spherical coordinates.

\begin{table}[!t]
\caption {Energetic flares observed in NOAA 12192 during October 22 - 25 2014} \label{tab:title}
\label{tab1}
\begin{tabular}{lccccccl} % define the column alignment
                                  % l: left, c: center, r: right
                                  % @{.} replace the inter-column by a .
  \hline
No. & \multicolumn{5}{c}{Flares (GOES)}& Location 
     & Nature of eruption  \\
 \cline{2-6}  
     &   Date & Start & Peak & End                  & Class 
     &    \\
  \hline
1 & 22 Oct. 2014 & 01:16 & 01:59 & 02:28 & M8.7 & S13E21  & Non eruptive  \\
2 & 22 Oct. 2014 & 14:02 & 14:06 & 22:30 & X1.6 & S14E13  & Non eruptive  \\
3 & 24 Oct. 2014 & 07:37 & 07:48 & 07:53 & M4.0 & S19W05  & Eruptive \\
4 & 24 Oct. 2014 & 20:50 & 21:15 & 00:14 & X3.1 & S16W21  & Non eruptive\\
5 & 25 Oct. 2014 & 16:55 & 17:08 & 18:11 & X1.0 & S10W22  & Non eruptive \\

  \hline
\end{tabular}
\end{table}

As the errors in the horizontal-field components of the vector magnetic field increase towards the limb, we have restricted our analysis to only those major flares produced by AR 12192 when the center position of the AR was well within $\pm$ 45$^{\circ}$ from the central meridian. Thus our data set contains three X-class flares and two M-class flares (Table 1) observed during 22 - 25 October 2014 close to disc center. The typical error in the transverse magnetic field of the SHARP data product is 100 Gauss, whereas the error in the line-of-sight component is 5 - 10 Gauss \citep{Liu,Hoeksema}.

As a complementary dataset we have used UV images (94 \AA, 1600 \AA) taken from SDO/AIA. AIA 94 \AA $ $ images represent the hot solar corona at a temperature of six million Kelvin. In particular, these images are useful to study the post flare loops filled with hot plasma due to chromospheric evaporation \citep{Doschek}. AIA 1600 \AA $ $ images represent the lower chromospheric region, which is known as the transition region of the solar atmosphere. These images are useful to identify the spatial locations of the flare ribbons, which are believed to be the low altitude impact of the accelerated particles flowing down from the reconnection site along the reconnected magnetic-field line (Priest and Forbes, 2002).

The sunspot group within AR 12192 belongs to the complex $\beta \gamma \delta$ class, with spots of opposite sign in the same penumbrae. AR 12192 showed morphological changes (Figure 2) as well as the changes in magnetic topology during its evolution throughout the disc passage. Therefore, the time-series analysis for each flare of our data set have been carried out within different bounded regions enclosing the locations of sheared horizontal field close to the polarity inversion line (PIL) where the major changes in horizontal field and Lorentz force are expected to occur \citep{Petrie}. In order to define the size, orientation, and location of the selected boxes we examined the post-flare loops in AIA 94 \AA $ $ images. As the post-flare arcades are believed to be the regions above which magnetic reconnection occurs \citep{Liuc,Chena,Joshi,Kumar}, we have selected our region of interest by enclosing the major post flare arcade structures seen in AIA 94 \AA $ $ images (Figure 3). A time-series analysis of magnetic-field evolution and Lorentz-force changes for the five major flares (Table 1) produced by AR 12192 was carried out on those bounded regions. To make sure that the selected boxes cover the region of sheared PIL where the major flares occurred, we have overplotted the flaring pixels observed in the AIA 1600 \AA $ $ image at the flare peak time and drawn the PIL on the magnetogram data (Figure 3) for each flare of our data set.
\begin{figure} [t]
\centerline{\includegraphics[width=12cm]{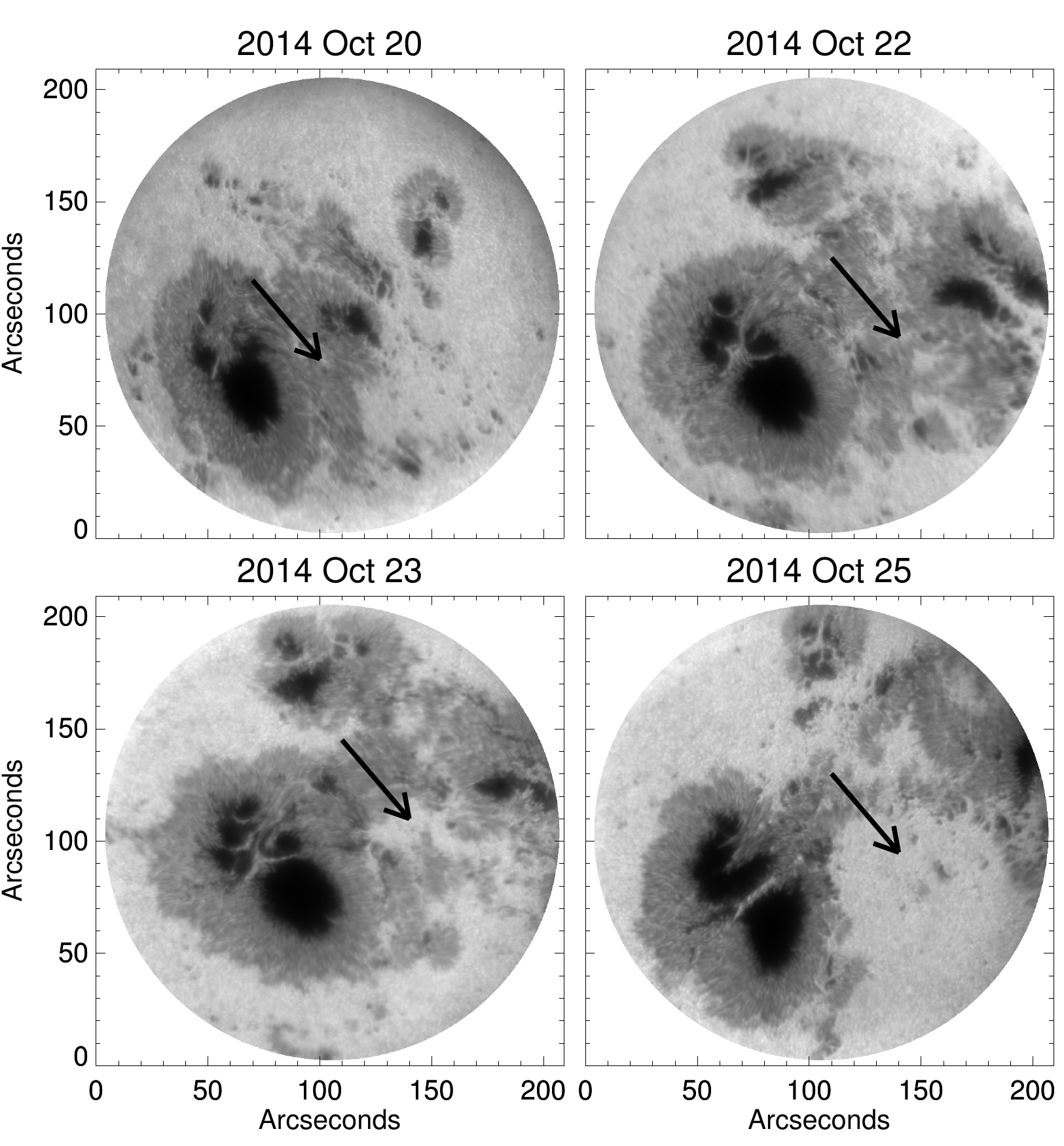}}
\caption{G-band images of AR 12192 taken from Multi Application Solar Telescope (MAST) at the Udaipur Solar Observatory. The black arrows mark the core region of AR 12192 where significant morphological changes took place. Due to the large size of AR 12192, the whole AR could not be captured within the field of view (three arcminutes) of MAST}%\label{fig:?}
\end{figure}
\begin{figure}[!t] 
\centerline{\includegraphics[trim={.5cm 0 .5cm 0},clip,width=15cm]{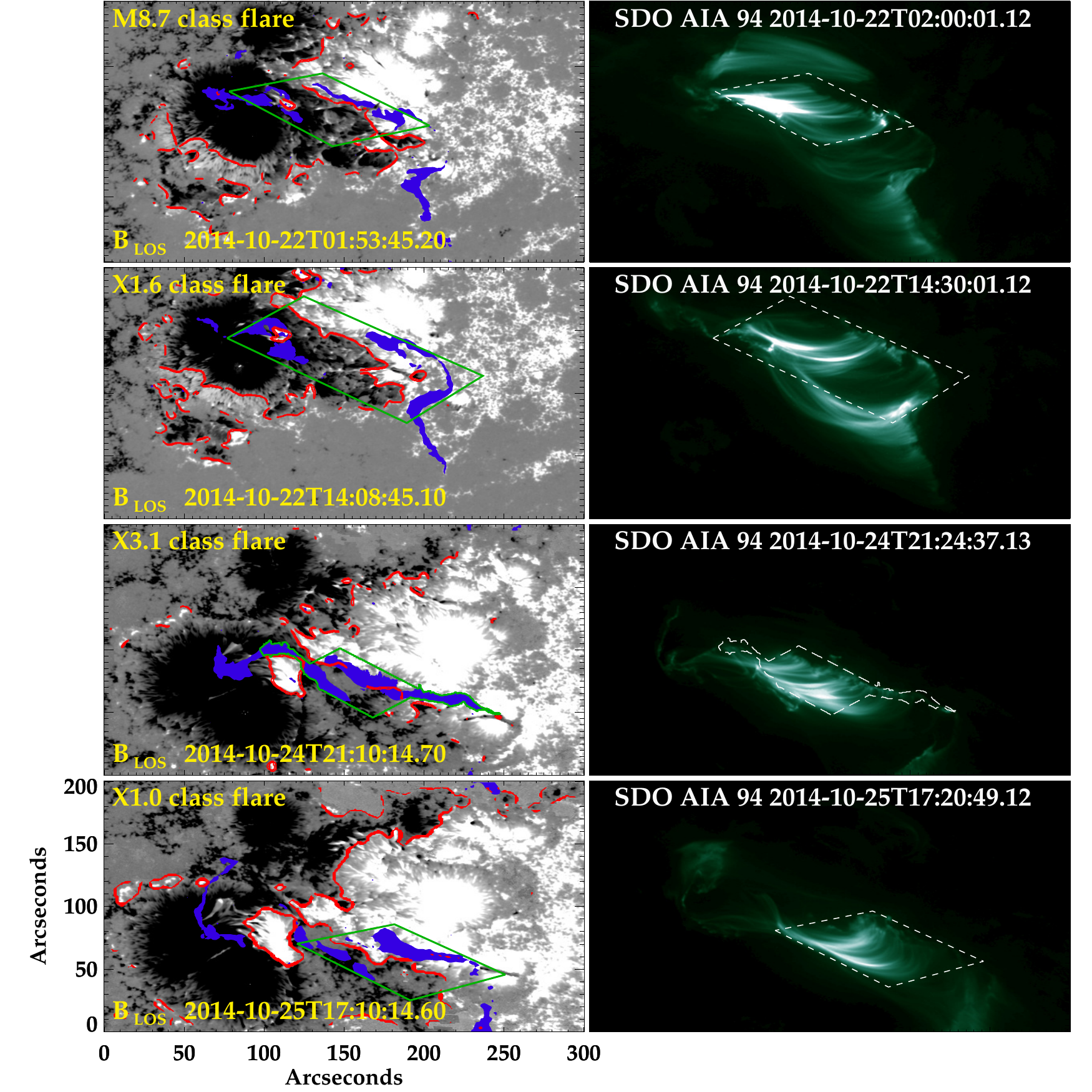}}
\caption{\textit{Left column}: HMI line-of-sight magnetic field during the non-eruptive flares. The red lines denote the polarity inversion line and the blue regions are the over-plotted flaring pixels from AIA 1600 \AA $ $ images. The green boundary denotes the selected region within which all the calculations have been done. \textit{Right column}: the post-flare arcade in AIA 94 \AA $ $ images. The white dashed boundary in each image of right column shows the selected region bounded by the green-solid line in the left column.  }%\label{fig:?}
\end{figure}

 For identification of the PIL we have followed the following steps: First we have convolved all of the pixels with a Gaussian kernel. Then, centering each pixel of the magnetogram of vertical component of magnetic field, we have scanned five consecutive pixels in both the horizontal and vertical directions. After that we compared the maximum and minimum values of each vertical and horizontal array of five pixels. If these two values are opposite in sign and the magnitude of these two values are greater than the noise level of ten Gauss on either of the two arrays, then the pixel that is in between these two minimum and maximum valued pixels is identified as the pixel tracing the PIL. These pixels were then joined to make the PIL.

 To calculate the Lorentz-force changes we have used the formulation introduced by \citet{Fisher}. The formulation is given as below,
    $$ \delta F_{\rm r}=\frac{1}{8\pi}\int_{A_{\rm ph}} (\delta {B_{\rm h}}^2-\delta {B_{\rm r}}^2)\\\ {\mathrm d}{\rm A} $$
    $$ \delta F_{\rm h}=-\frac{1}{4\pi}\int_{A_{\rm ph}}\delta ({B_{\rm h}}{B_{\rm r}})\\\ {\mathrm d}{\rm A} $$
   where $ B_{\rm h} $ and $ B_{\rm r} $ are the horizontal and radial components of the magnetic field, $F_{\rm h}$ and $F_{\rm r}$ are the horizontal and radial components of the Lorentz force calculated over the volume of the active region, $\rm A_{\rm ph} $ is the area of the photospheric domain containing the active region, and dA is the elementary surface area on the photosphere.

To study the flare-related morphological changes of AR 12192 we have used the full-disc continuum images observed in the Fe I absorption line at 6173 \AA $ $ with a spatial scale of 0.5$''$ per pixel and temporal scale of 12 minutes. We have taken a cutout of the whole active region from the full-disc continuum intensity map and then all of the images were differentially rotated to the solar disc center. To identify the umbra--penumbra and penumbra--quiet-Sun boundaries, we first have normalized the brightness values of all the pixels within the cutout region by the median of brightness values of a 10$\times$10 pixel$^2 $ quiet-Sun region surrounding the sunspot. Then a cumulative histogram \citep{Pettauer,Mathew} of the intensity of each pixel's brightness is computed within the cutout, which encloses the umbral and penumbral region as well as the immediately surrounding quiet-Sun region. This cumulative histogram (Figure 4) is used to calculate the intensity threshold for defining the umbra--penumbra and penumbra--quiet-Sun boundary. 

\begin{figure}[!t] 
\centerline{\includegraphics[width=10cm]{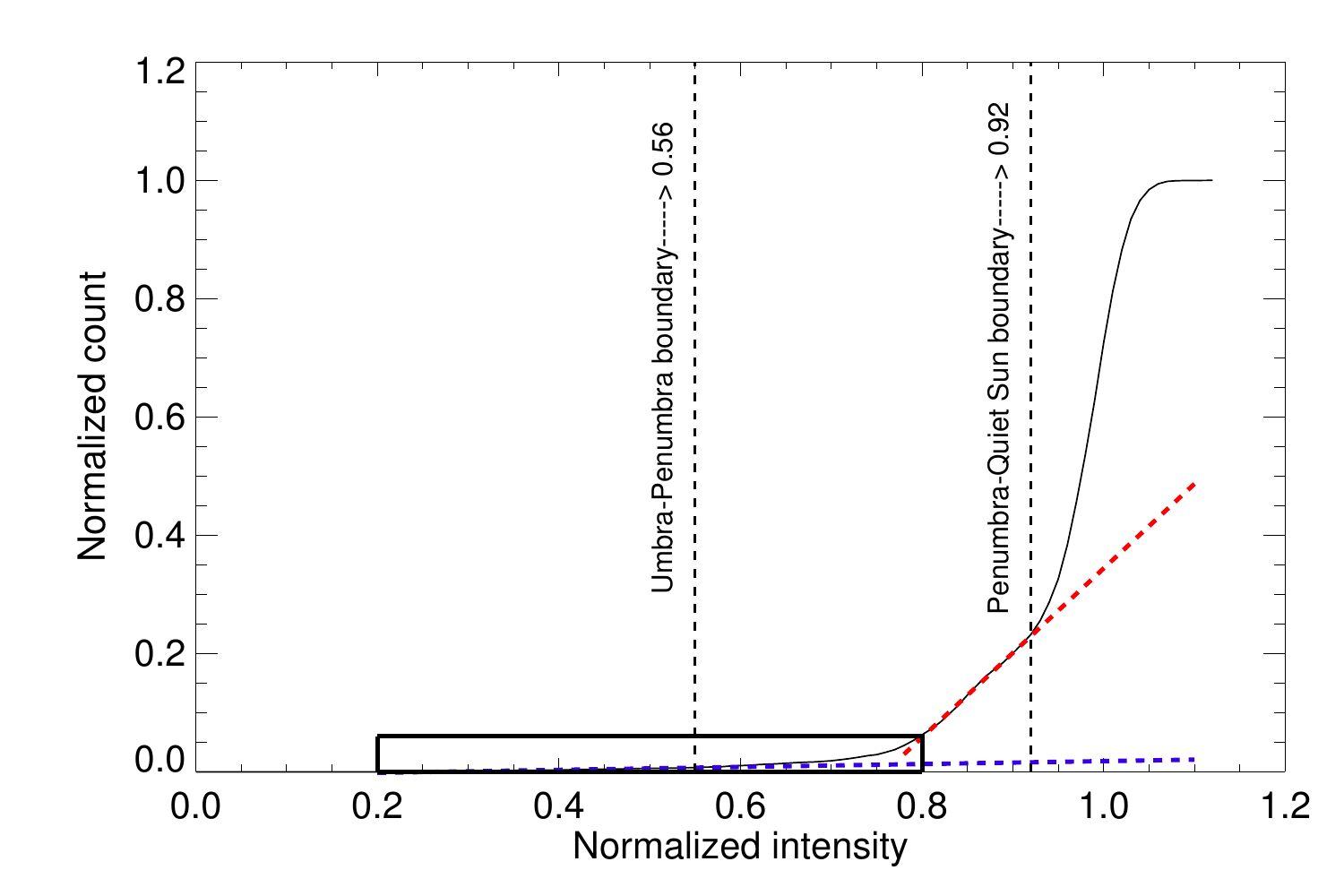}}
\centerline{\includegraphics[width=10cm]{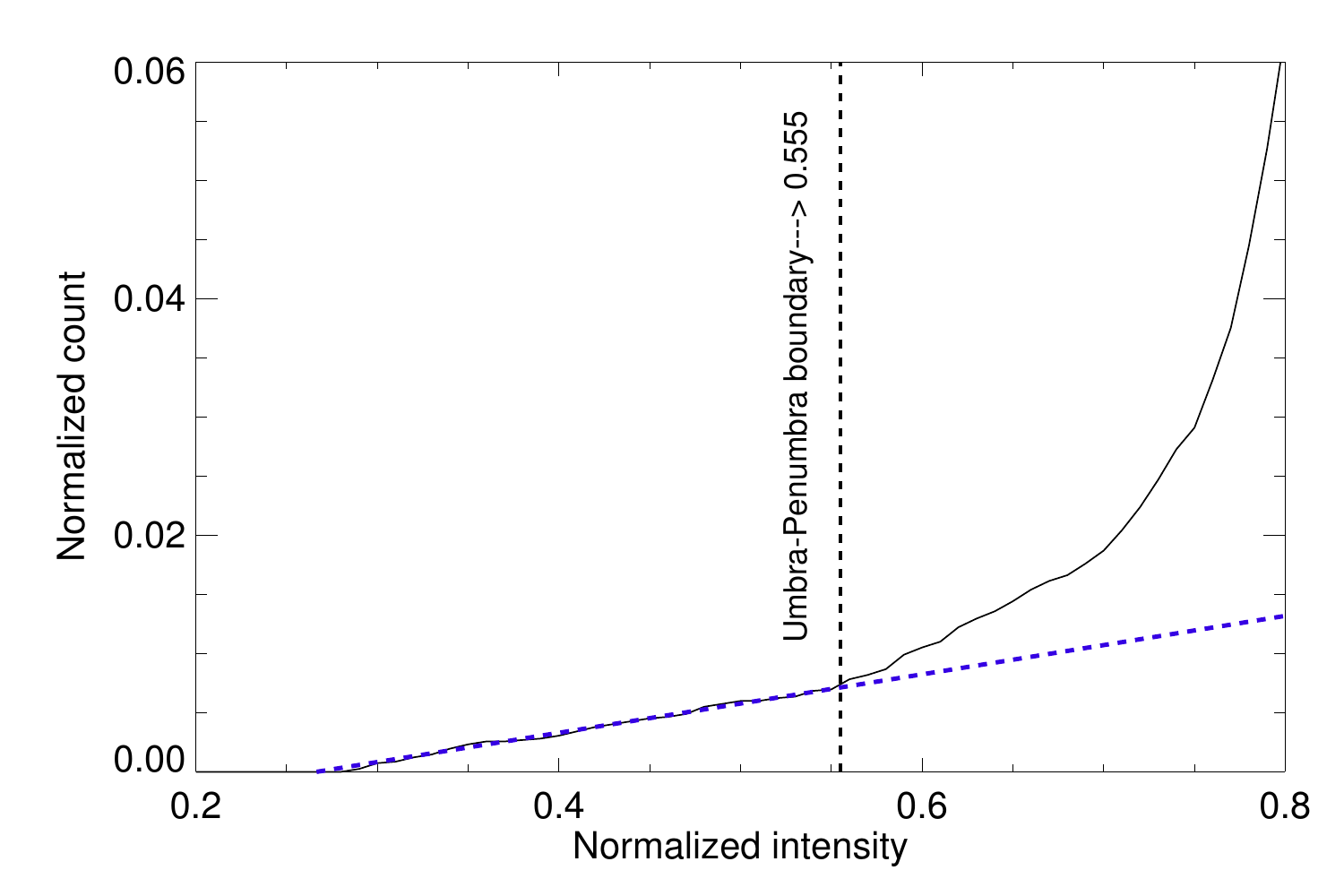}}
\caption{\textit{Top panel}: the cumulative intensity histogram for identifying the umbra--penumbra and penumbra--quiet-Sun boundary. The blue and red dashed lines in the top panel are the linearly fitted lines to the two nearly flat part of the plot corresponding to umbral and penumbral region respectively. The vertical dashed lines mark the values obtained for umbra--penumbra and penumbra--quiet-Sun boundaries. \textit{Bottom panel}: the linearly fitted flattest part of the plot enlarged from the rectangular box shown in the top panel (within normalized intensity range 0.2 to 0.8 and normalized count values 0.00 to 0.06).}
\end{figure}

The steep rise in the histogram plot (upper panel of Figure 4) near the normalized intensity unity corresponds to the quiet-Sun region, whereas the flattest part of the plot corresponds to the umbral region. In between the above two parts of the plot, the less steep and moderately flatter part denotes the penumbral region. In order to get the threshold values, both of the nearly flat parts of the plot were linearly fitted. The maximum intensity at which the linearly fitted straight lines start deviating from the plot is considered to be the intensity threshold. Applying the above procedure to the cumulative histogram computed from the cutout region enclosing the source location of eruptive M4.0-class flare on 24 October 2014, we have found the normalized intensity threshold for the umbra--penumbra boundary is 0.56 and that of penumbra--quiet-Sun boundary is 0.92. These values were then used to calculate the penumbral area variation during the eruptive M4.0-class flare. To study the overall umbral and penumbral area variation of the whole AR during its disc passage, we have used the normalized intensity threshold for the umbral--penumbral boundary as 0.53 and that of the penumbral--quiet-Sun boundary as 0.90 \citep{Sarkar}. In order to remove the projection effect of the spot area, we have used the algorithm introduced by \citet{Cakmak}.

 Furthermore, to investigate the overlying magnetic-field strength for both the eruptive and non-eruptive region of AR 12192, we have extrapolated the photospheric magnetic field using the non linear force free field (NLFFF) model \citep{Aschwanden}. From the extrapolated magnetic field we have calculated the decay index over the non-eruptive core region and the eruptive flaring region of AR 12192. The decay index [n] is defined as $$n= - \frac{\partial {\rm log}(B_{\rm ex})}{\partial {\rm log}(h)} \ ,$$ where h is the height above the solar photosphere and $B_{\rm ex}$ is the external magnetic field obtained from the extrapolated field. According to the definition, as the decay index [n] quantifies how the external magnetic field is decaying with respect to height, it has been used to infer the constraint of overlying magnetic-field strength in several studies \citep{TorokKliem,WangZang,Liu2008}. The critical decay index for onset of the torus instability has been found to be $\approx$1.5 \citep{Tor, Aulanier}. 
\section{Results}
\subsection{Magnetic-Field Evolution for the Non-Eruptive Flares}
 \begin{figure} [h]
\centering
  \includegraphics[trim={0.4cm 1cm 1cm .3cm},clip,width=.49\textwidth]{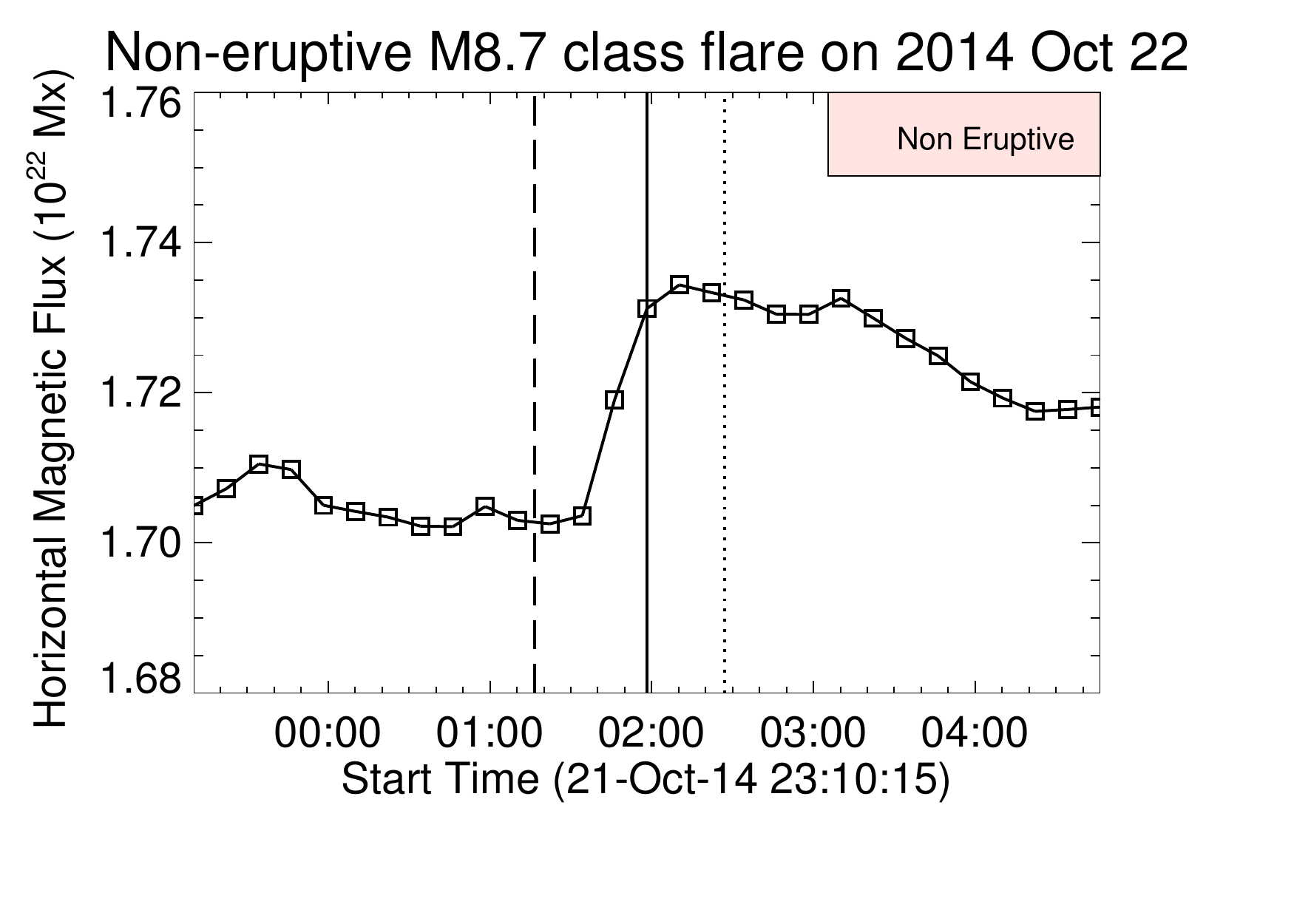}
  \centering
  \includegraphics[trim={0.4cm 1cm 1cm .3cm},clip,width=.49\textwidth]{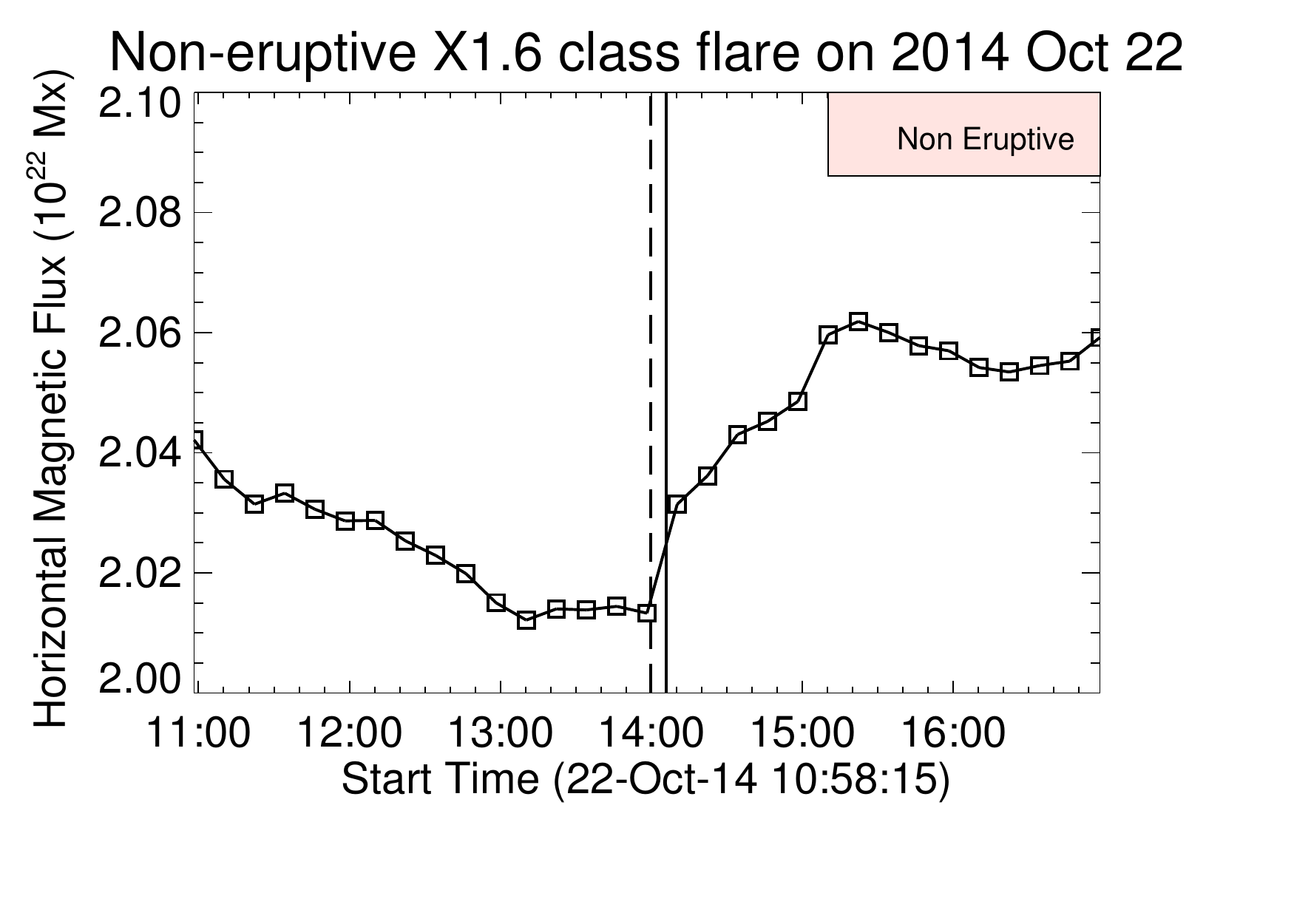}
  \centering
  \includegraphics[trim={0.4cm 1cm 1cm .3cm},clip,width=.49\textwidth]{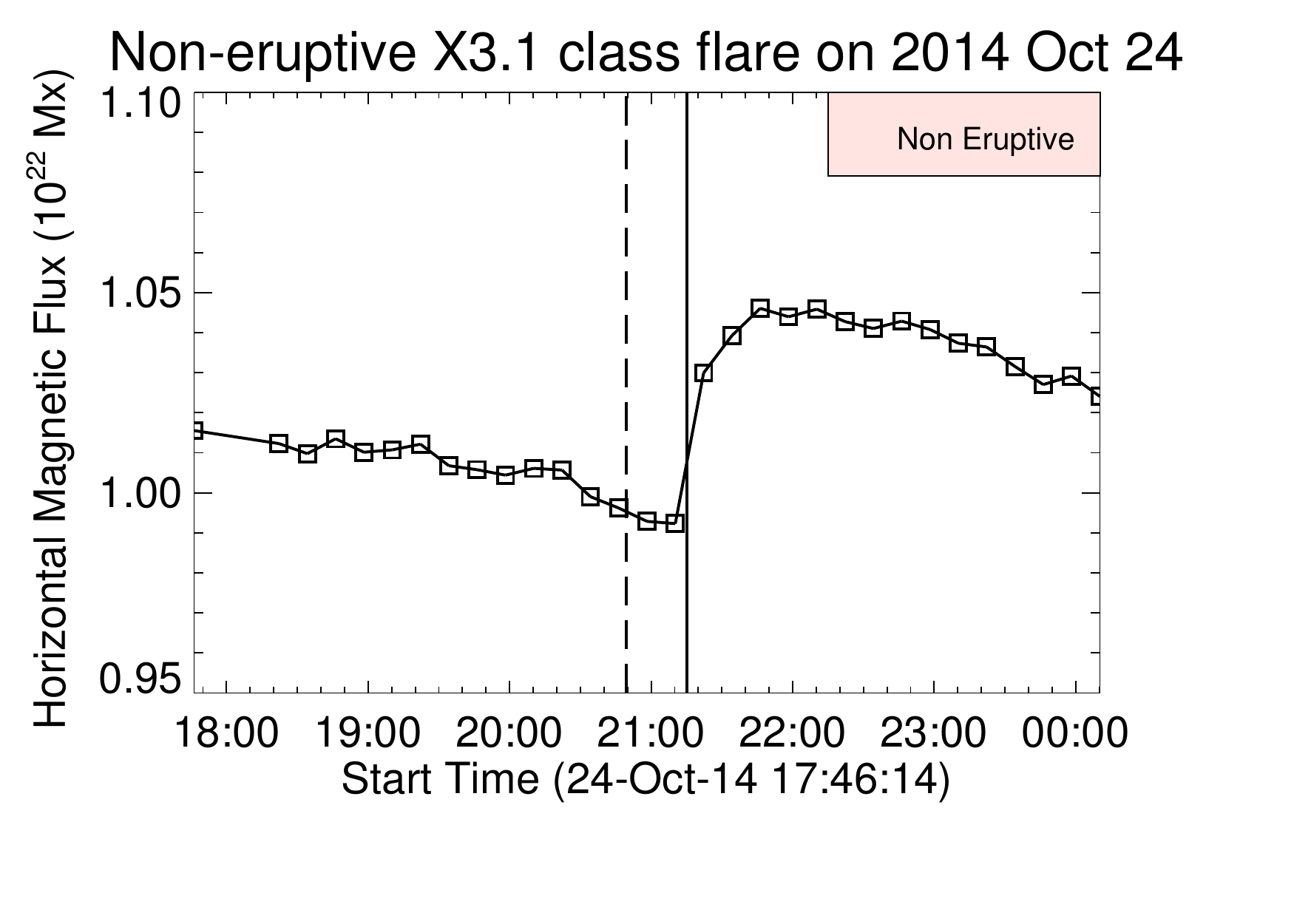}
  \centering
  \includegraphics[trim={0.4cm 1cm 1cm .3cm},clip,width=.49\textwidth]{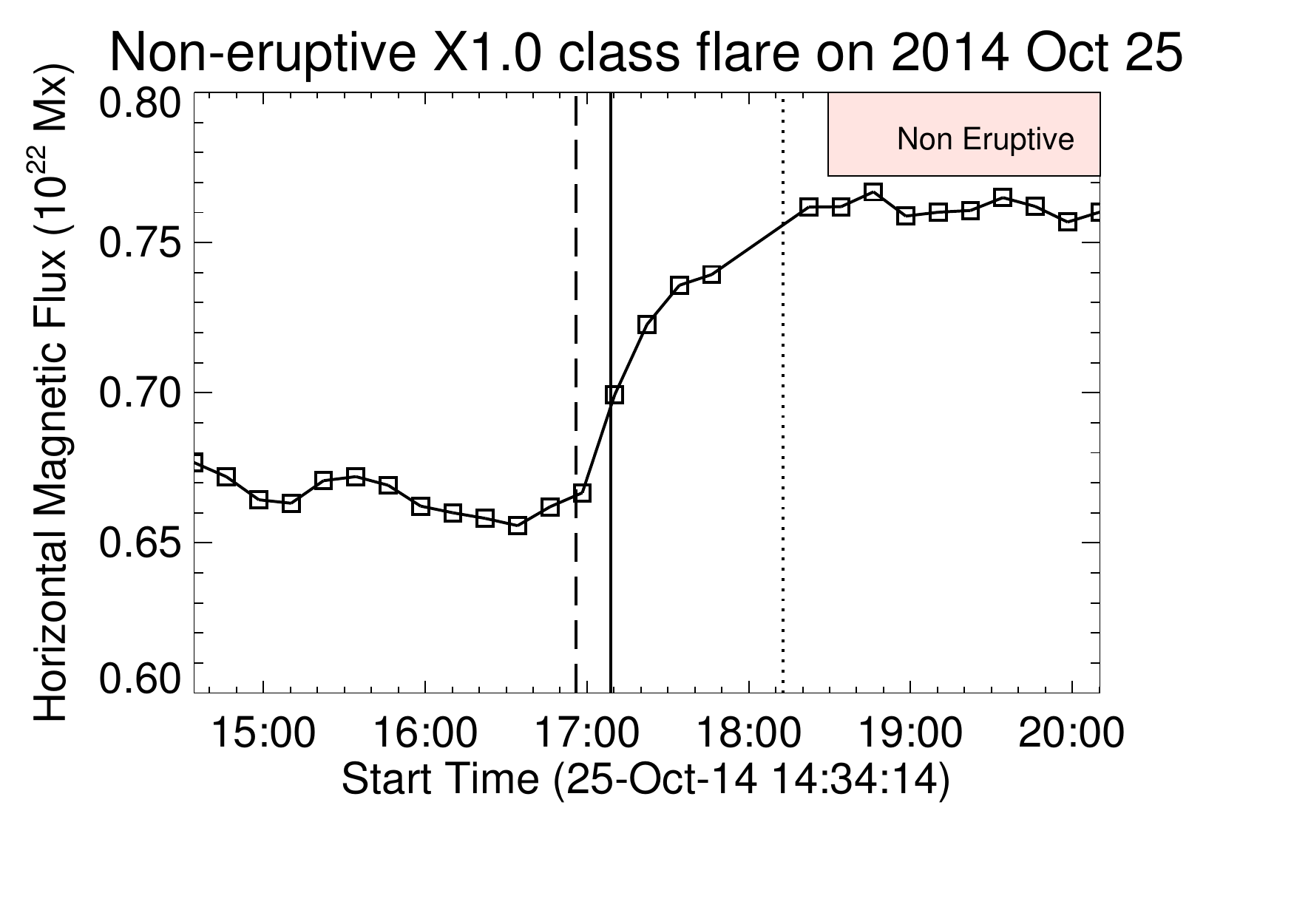}
\caption{Horizontal magnetic-field evolution during the non-eruptive flares. The dashed, solid, and dotted vertical lines denote the flare onset, peak, and decay times respectively.}%\label{fig:?}
\end{figure}
\pagebreak

Analyzing the six-hour series of vector magnetogram data for each of the non-eruptive flare of our dataset, we have found that the integrated transverse magnetic flux over the selected regions near the PIL, increased permanently for all of the cases (Figure 5). This result is consistent with the earlier flare-related  transverse-magnetic-field changes reported by \citet{Wang} and Petrie (2012). Here the changes in the horizontal magnetic flux have been found to be persistent up to more than two hours after the peak time of each flare. These permanent changes ensure that the temporal evolution of the horizontal magnetic flux was real and did not include any flare-related artifact, as the flare-related artifacts are transient in nature and do not cause any permanent changes in the temporal profile of the measured magnetic field \citep{Sun}. Figure 5 shows the temporal evolution of the horizontal magnetic flux for the non-eruptive flares. Error bars in the horizontal magnetic flux were too small ($\approx10^{-3}$ times the calculated values) to be plotted. The change in horizontal magnetic flux has been found to range between 3$\times$10$^{20}$ and 7$\times$10$^{20}$ Mx within less than half an hour for the four non-eruptive flares. In percentage terms, these changes range from about 2 \% to 10 \% from its initial value in the pre-flare stage.\\      
  \begin{figure} [!h]
\centering
  \includegraphics[trim={0.4cm 1cm .6cm .1cm},clip,width=.49\textwidth]{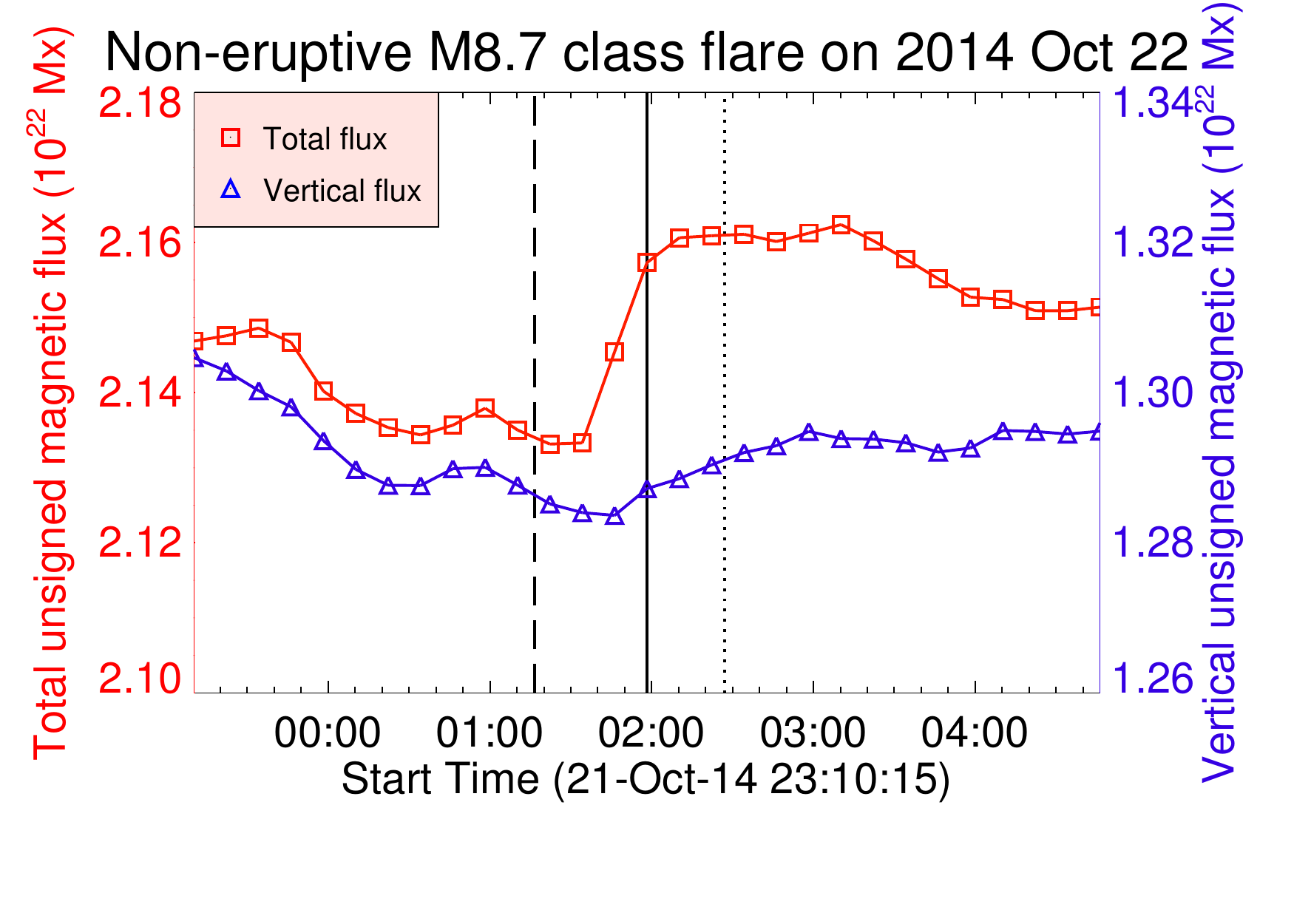}
  \centering
  \includegraphics[trim={0.4cm 1cm .6cm .1cm},clip,width=.49\textwidth]{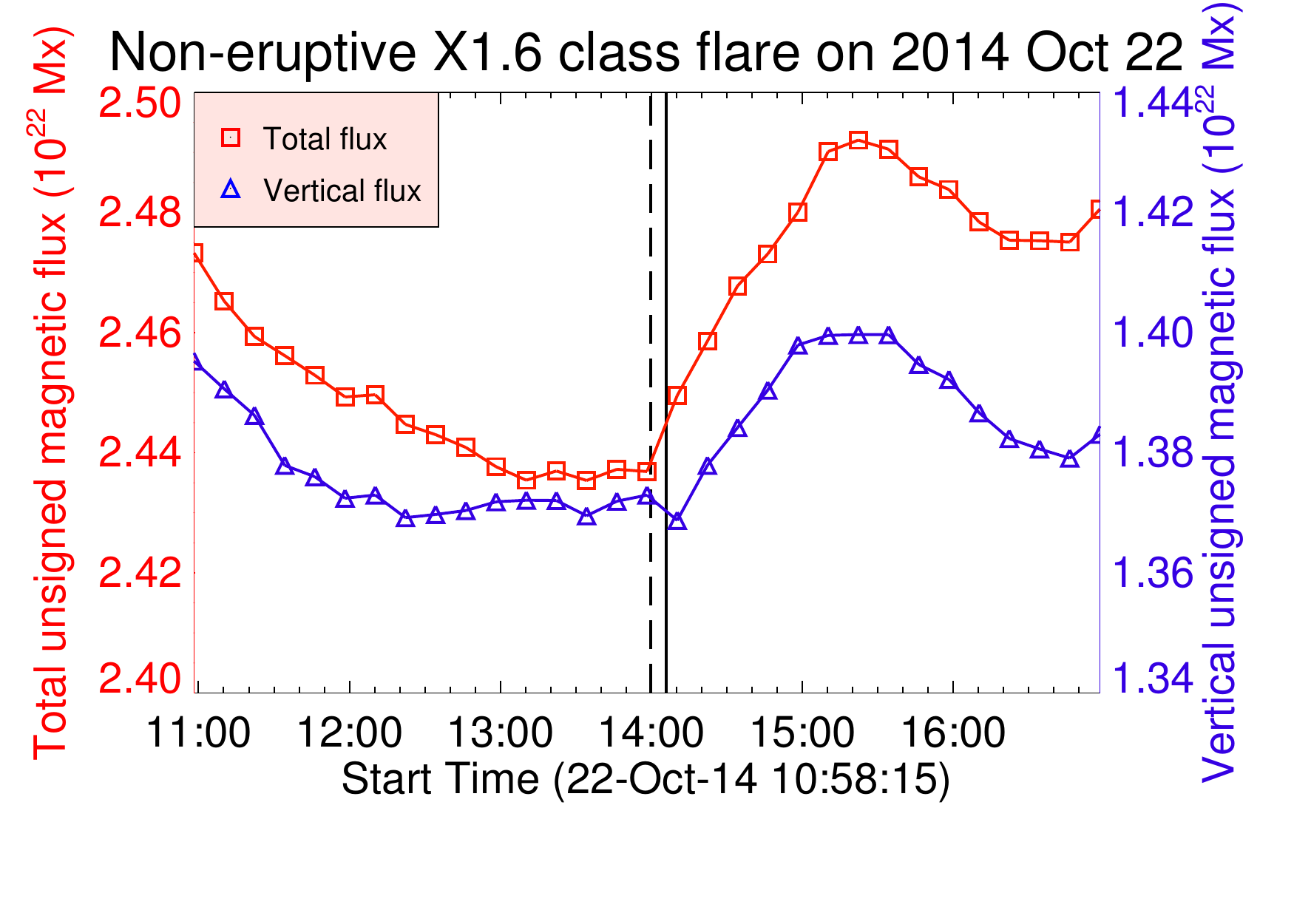}
  \centering
  \includegraphics[trim={0.4cm 1cm .6cm .1cm},clip,width=.49\textwidth]{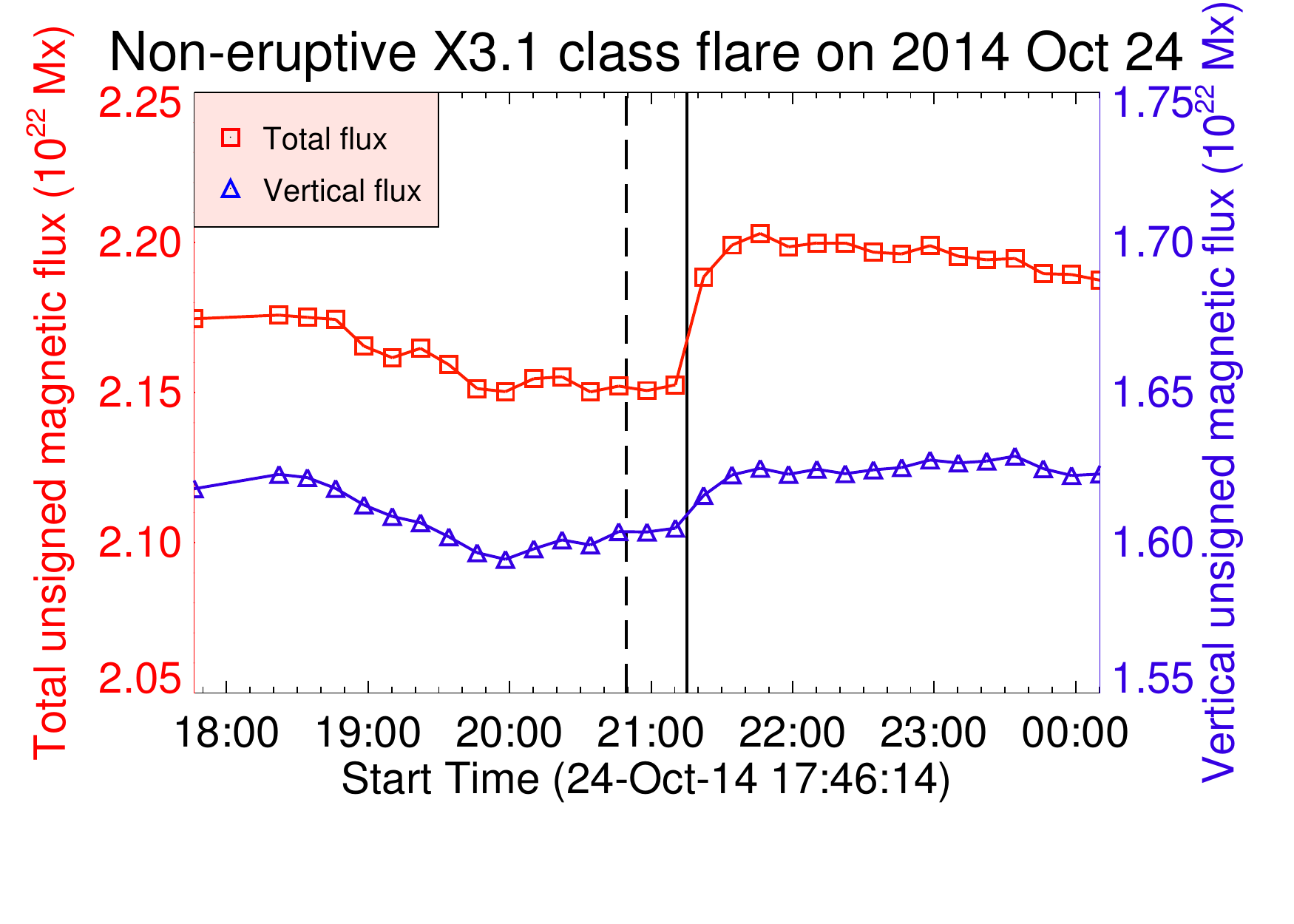}
  \centering
  \includegraphics[trim={0.4cm 1cm .6cm .1cm},clip,width=.49\textwidth]{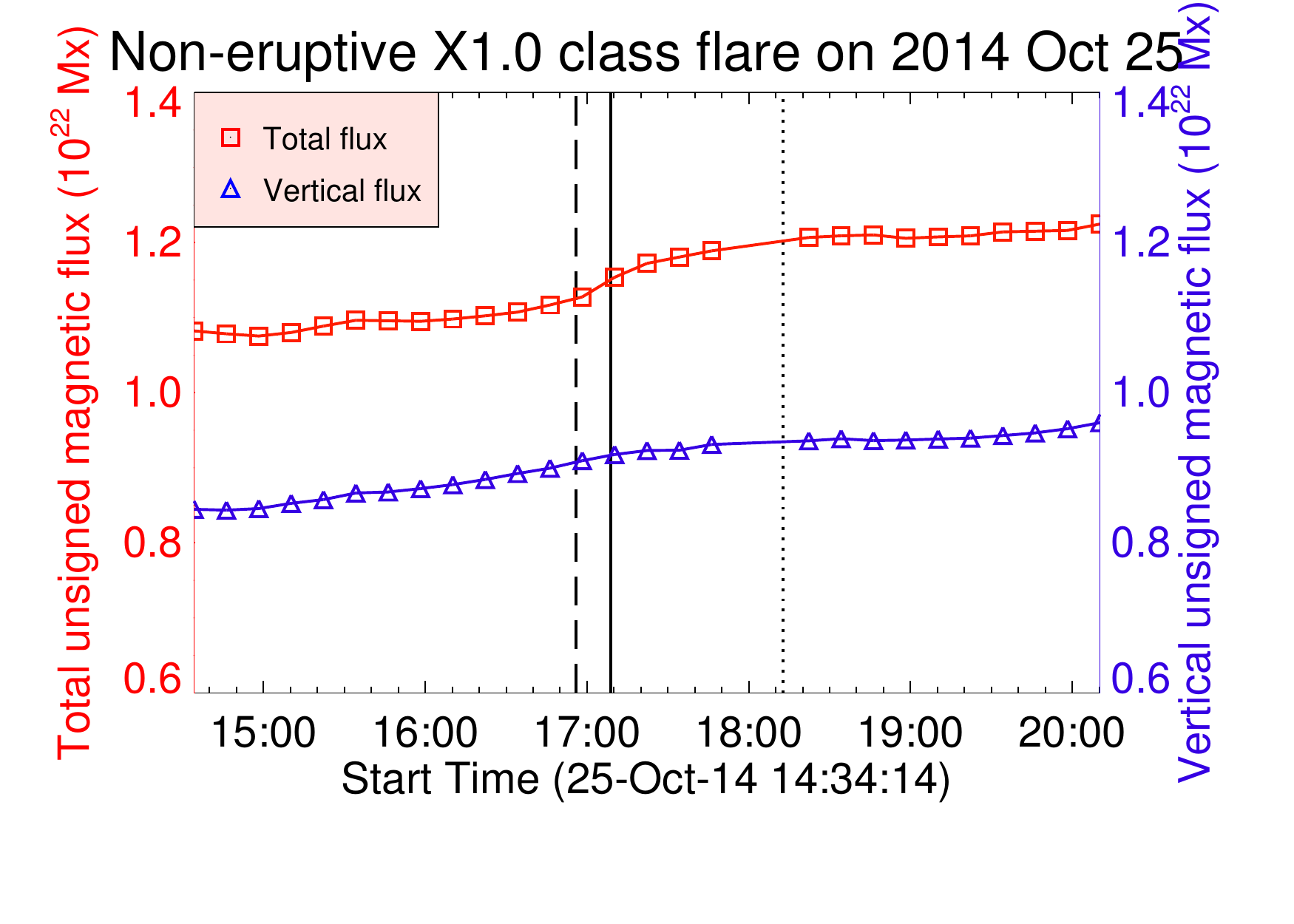}
  \caption{Temporal evolution of the total and vertical magnetic flux for the non-eruptive flares. The dashed, solid, and dotted vertical lines denote the flare onset, peak, and decay times respectively.}%\label{fig:?}
\end{figure}

 In Figure 6, we show the temporal evolution of the total and vertical magnetic flux. The red-solid lines in the figure depict the changes in total magnetic flux during the non-eruptive flares. Notably, the temporal profile of the vertical magnetic flux does not show any significant changes during the flares, which is in agreement with earlier results (Petrie 2012). The temporal profiles of the positive and negative flux (Figure 7) show that there was no significant flux cancellation along the PIL associated with the confined flares. Importantly, the absence of photospheric-flux cancellation underlying the post-flare loops associated with the non-eruptive flares confirms that there were no converging motions of the foot points of opposite magnetic polarities which can result in photospheric foot-point reconnection along the PIL to form the flux rope. These observations support the idea of tether-cutting reconnection reported by \citet{Chen2015} and \citet{Jiang}. They considered the shearing motion of the photospheric fluxes to be responsible for stressing  the coronal magnetic field and building  up  a large current sheet that triggered the magnetic reconnection and produced the homologous high energetic flares.

\begin{figure} [!t]
\centering
  \includegraphics[trim={0 1cm 1cm .3cm},clip,width=.49\textwidth]{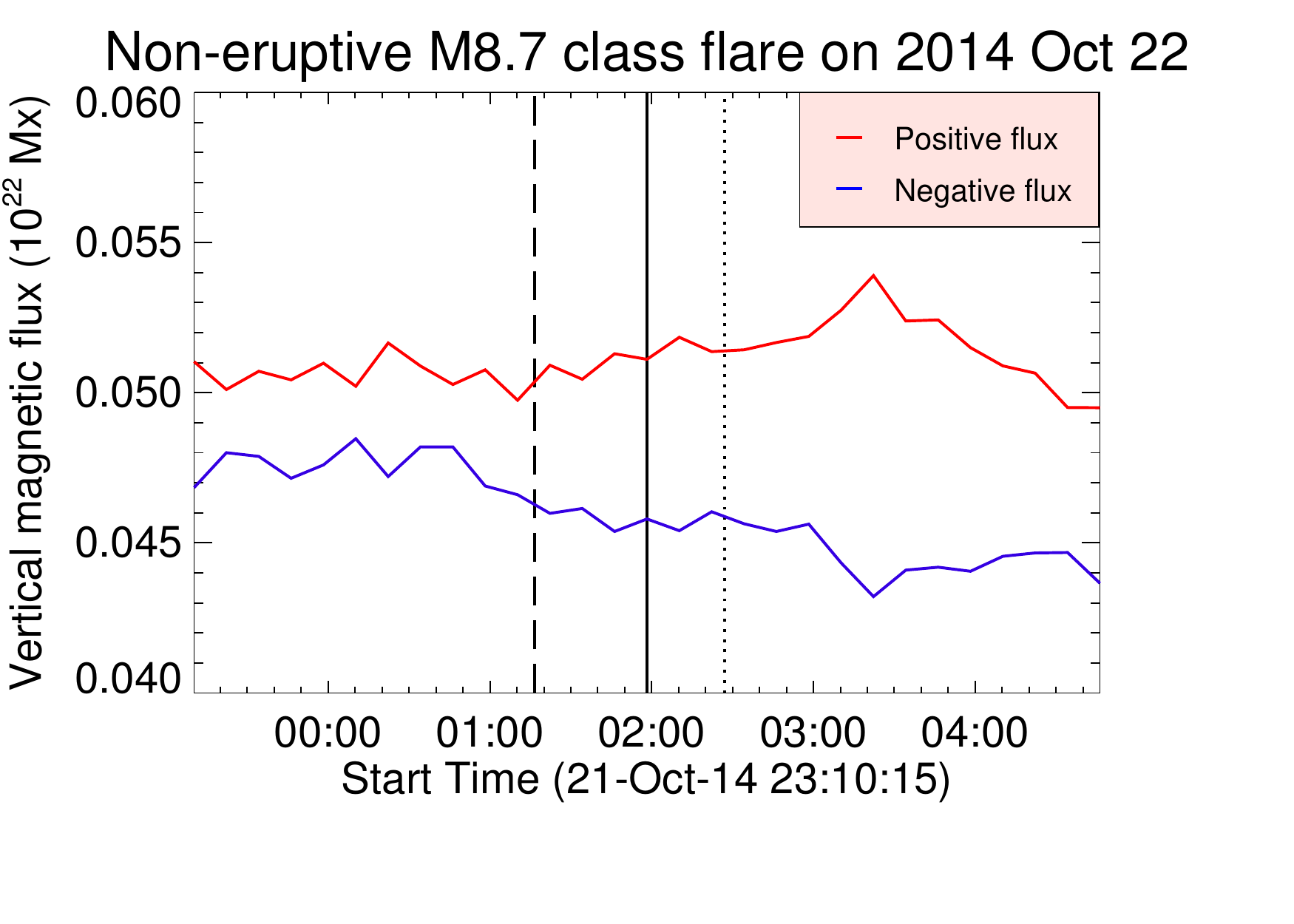}
  \centering
  \includegraphics[trim={0.2cm 1cm 1cm .3cm},clip,width=.49\textwidth]{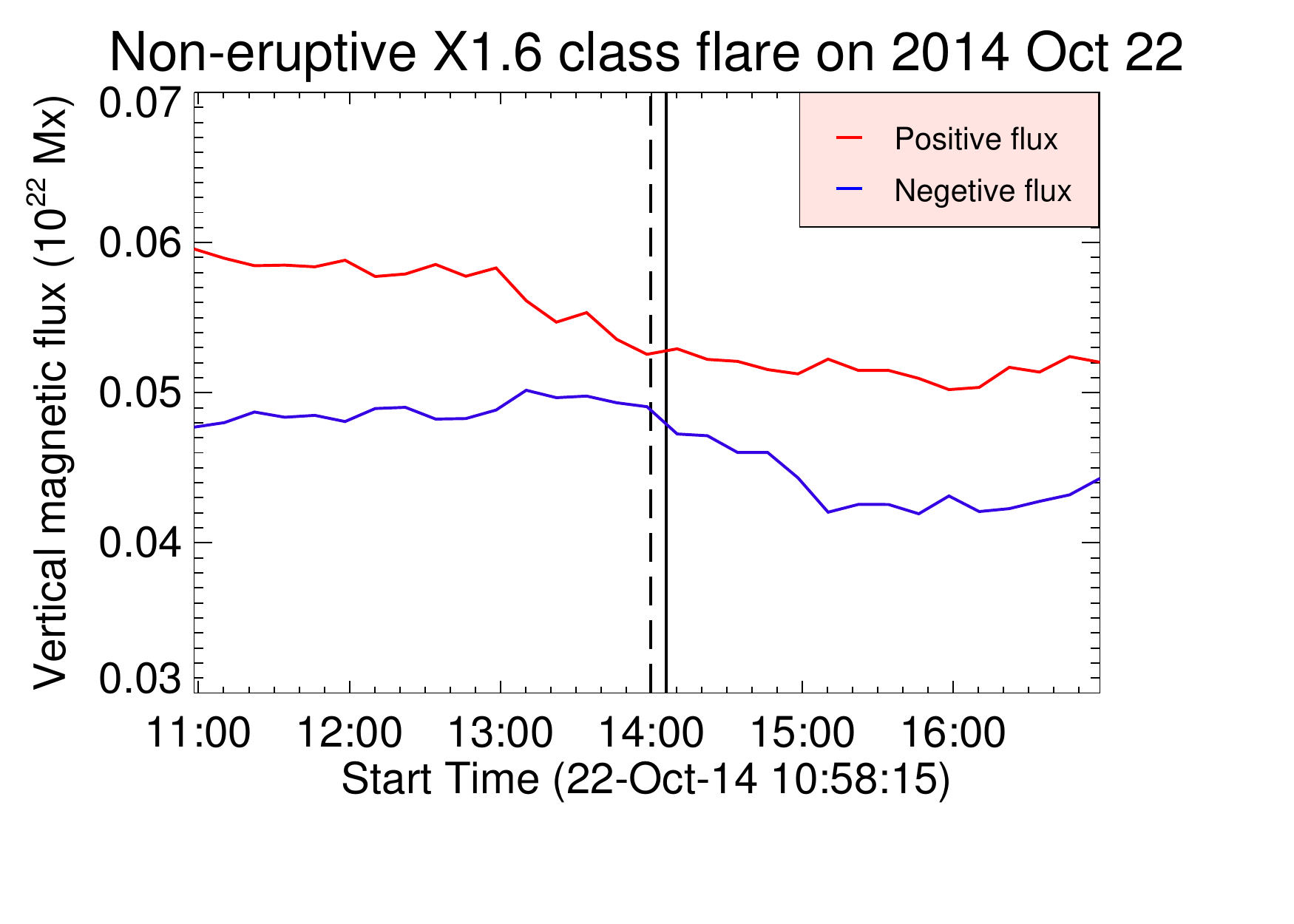}
  \centering
  \includegraphics[trim={0.4cm 1cm 1cm .3cm},clip,width=.49\textwidth]{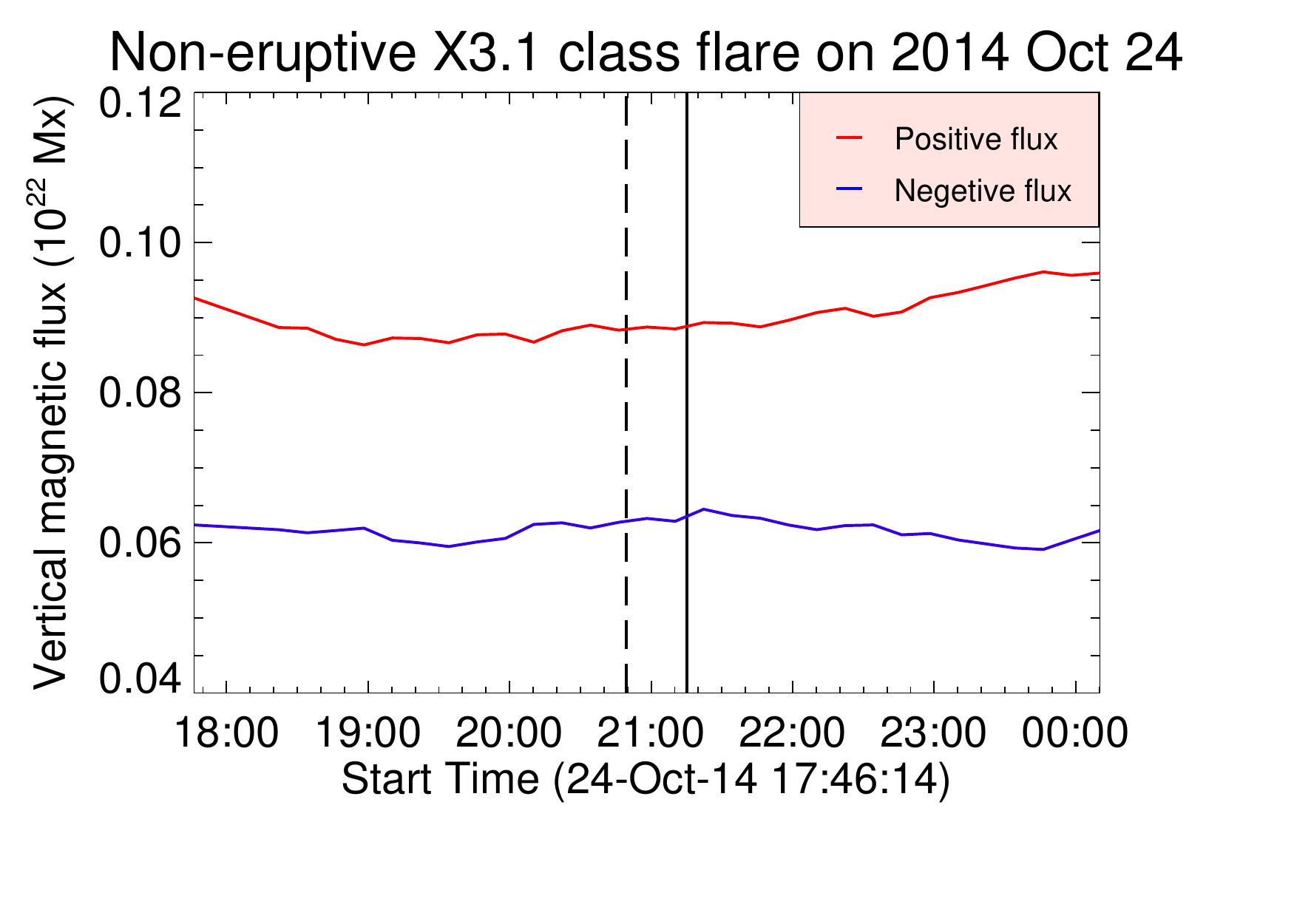}
  \centering
  \includegraphics[trim={0.2cm 1cm 1cm .3cm},clip,width=.49\textwidth]{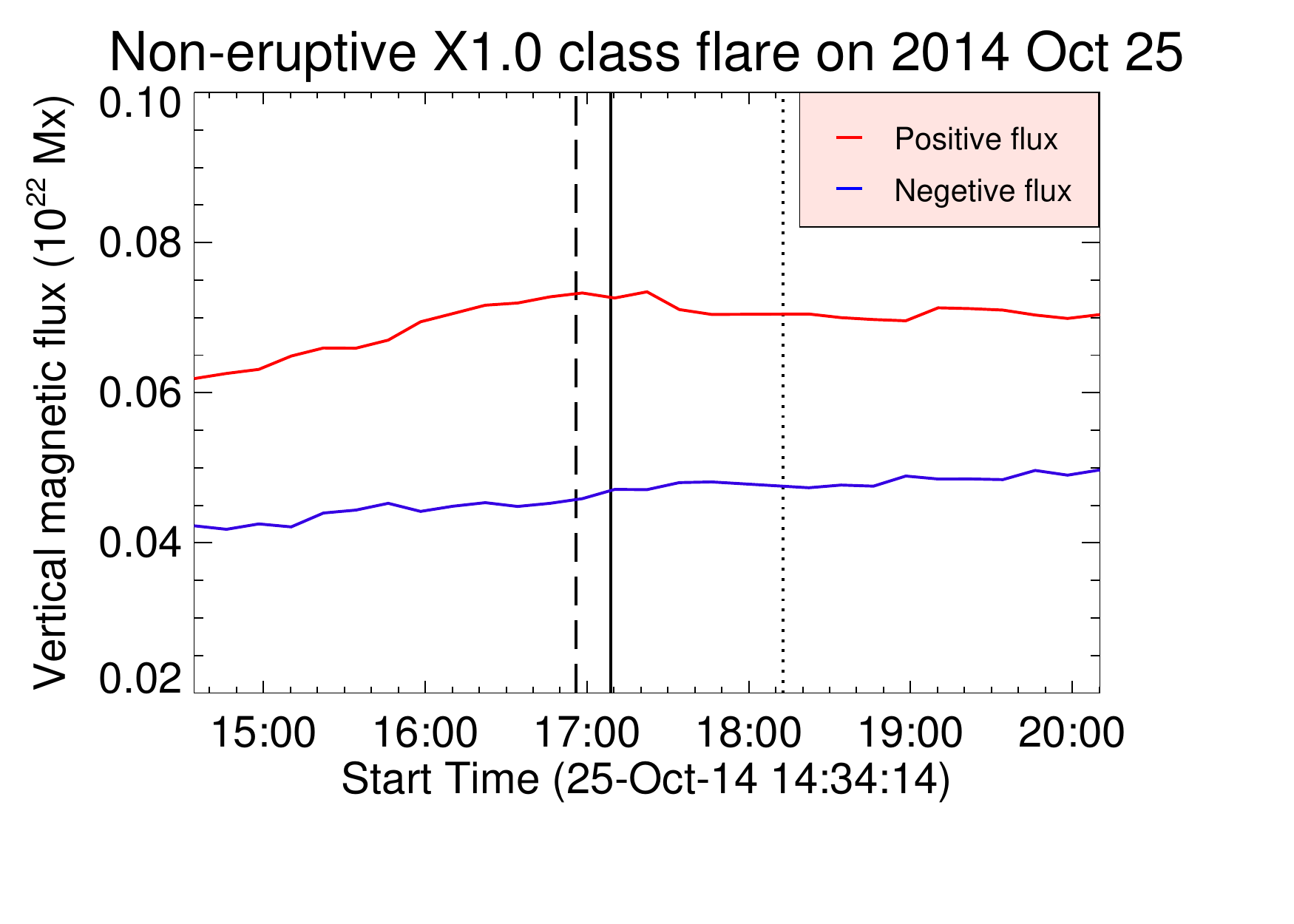}
\caption{The temporal evolution of positive and negative magnetic flux for the non-eruptive flares. The dashed, solid, and dotted vertical lines denote the flare onset, peak, and decay times respectively.}%\label{fig:?}
\end{figure}
 
The Lorentz-force changes for the non-eruptive flares have been illustrated in Figure 8. During each of the flares, the radial component of the Lorentz force underwent a large and abrupt downward change. The magnitude of these force changes, ranging from about 2.0$\times$10$^{22}$ dyne to 2.6 $\times$ 10$^{22}$ dyne, are comparable to those found in the previous estimates of flare-related Lorentz-force changes \citep{Petrie,ShouWang}. For the four non-eruptive flares, the change in Lorentz force per unit area ranges from about 390 dyne cm$^{-2}$ to 1390 dyne cm$^{-2}$. 
\begin{figure} [!t]
\centering
  \includegraphics[trim={0.7cm 1cm .8cm .25cm},width=.49\textwidth]{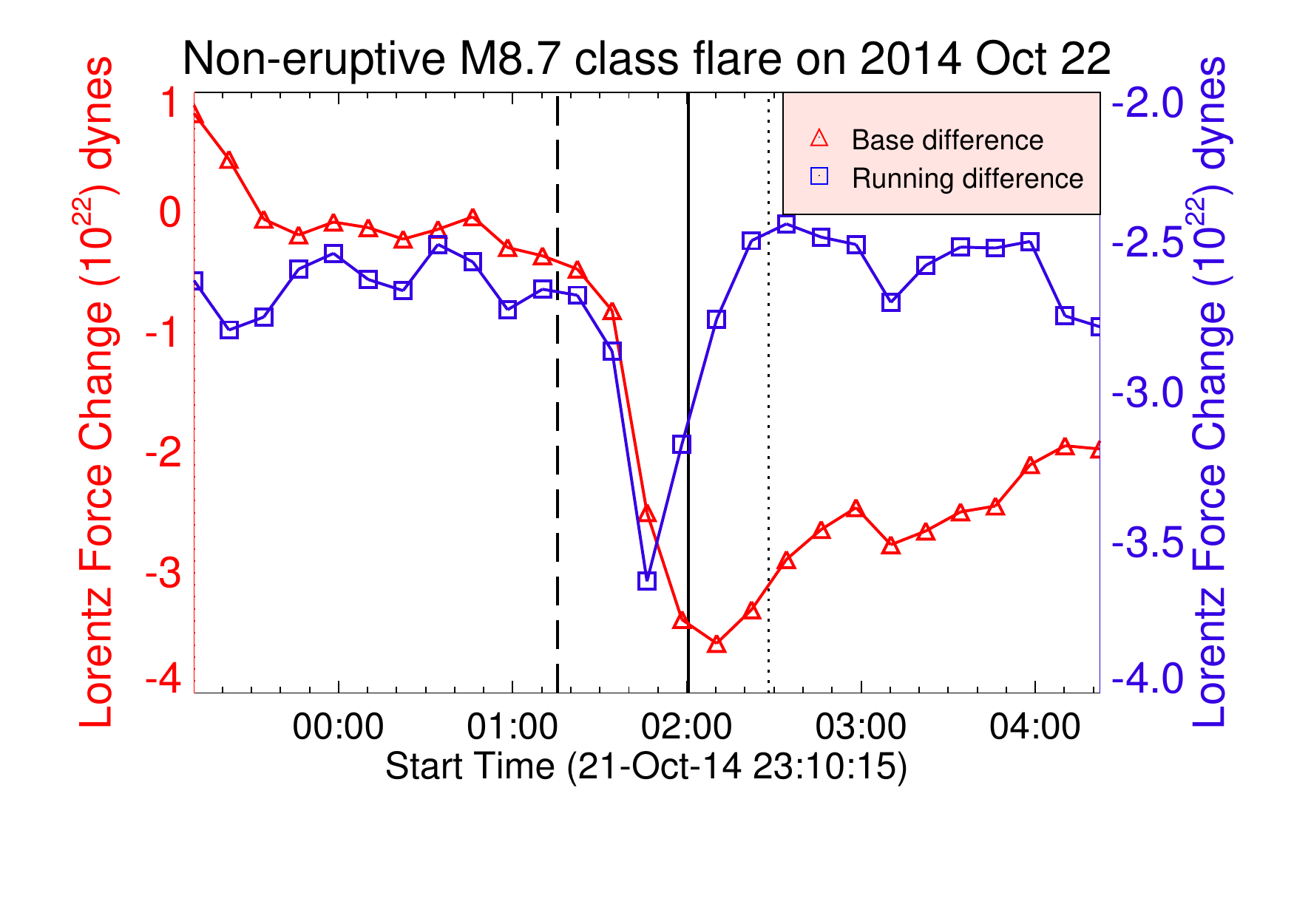}
  \centering
  \includegraphics[trim={0.3cm .9cm 1cm .65cm},width=.49\textwidth]{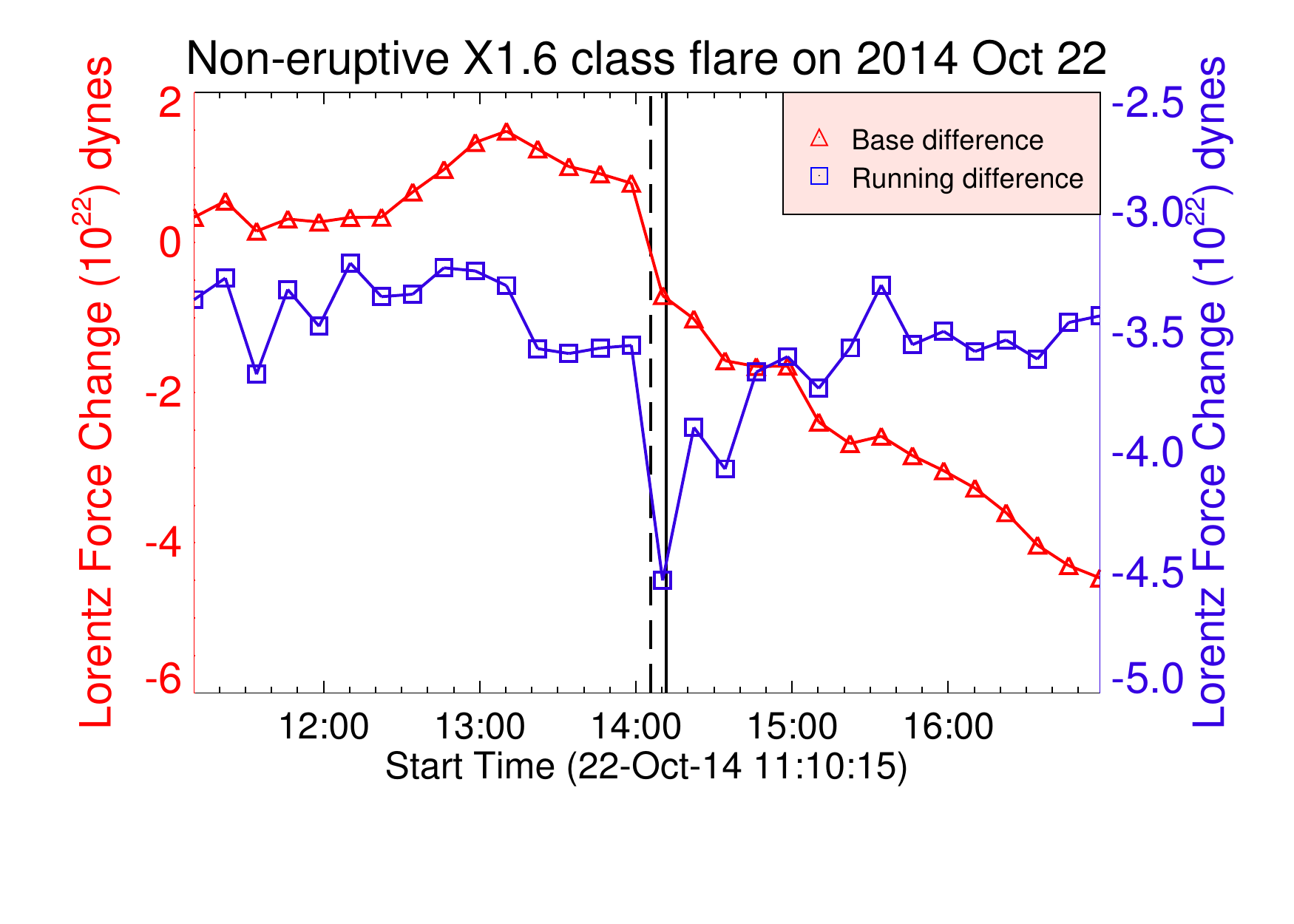}
  \centering
  \includegraphics[trim={0.8cm 1cm 1cm .3cm},width=.49\textwidth]{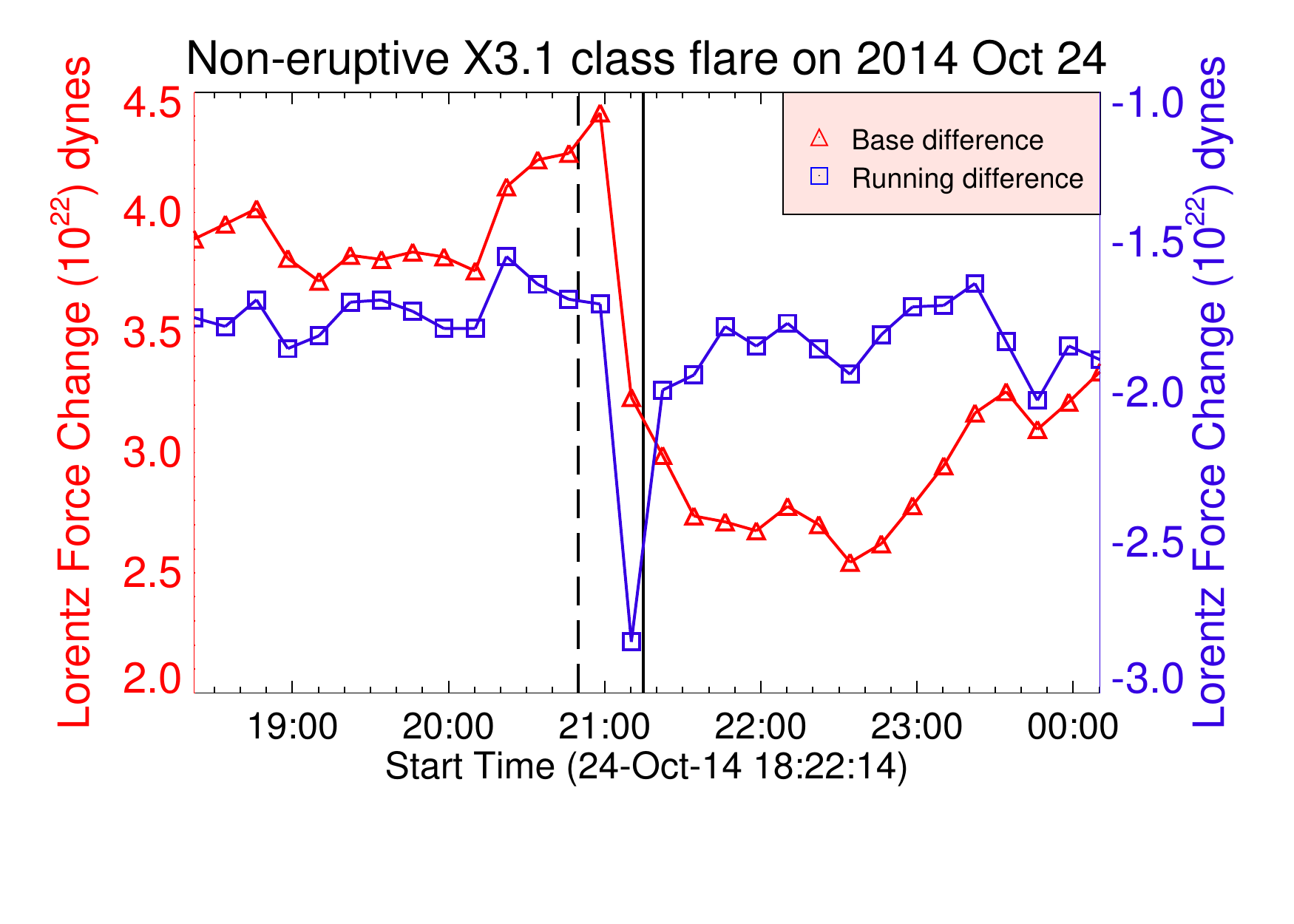}
  \centering
  \includegraphics[trim={0.6cm 1cm 1cm .65cm},width=.49\textwidth]{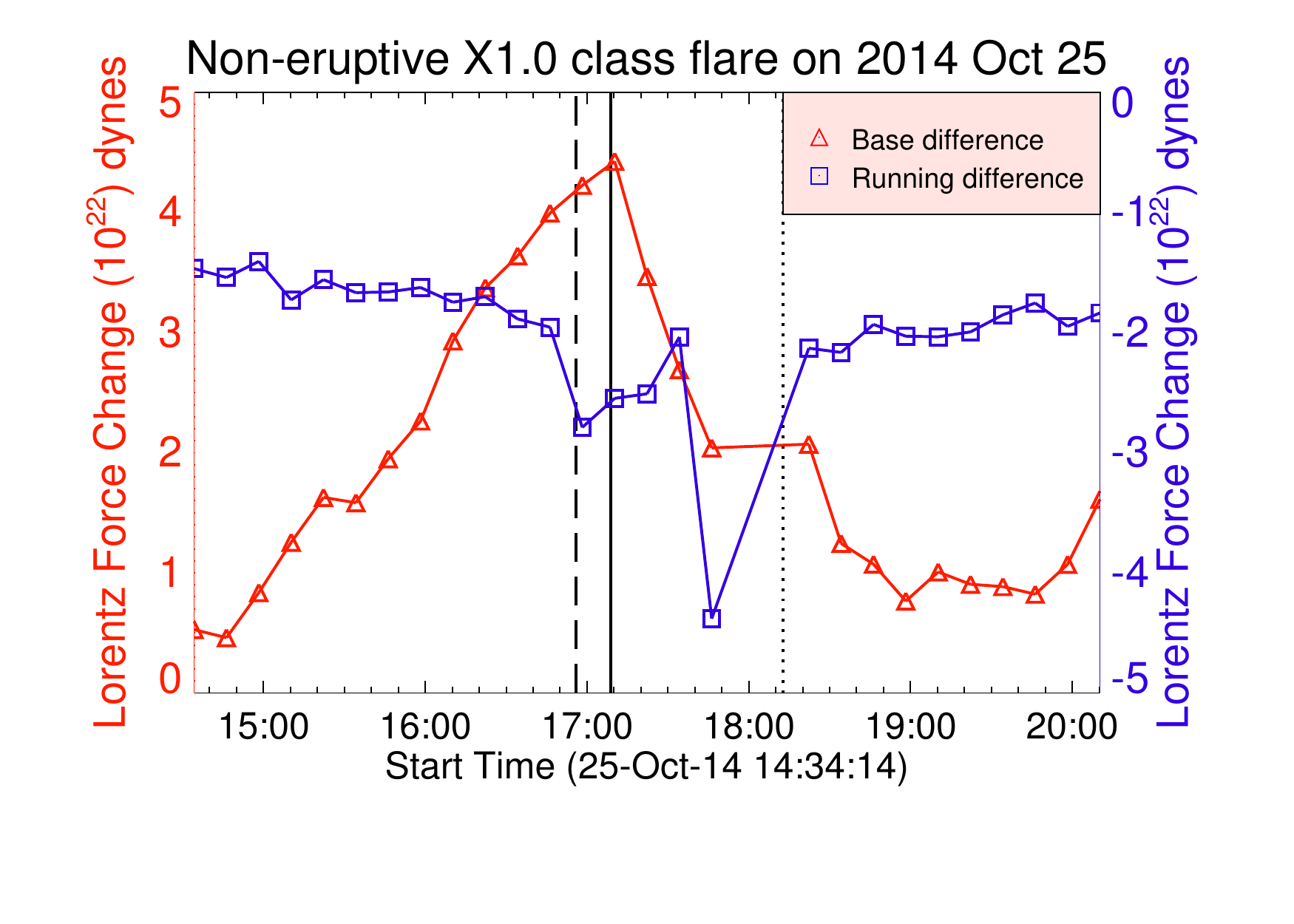}
\caption{The changes in radial component of Lorentz force for the non-eruptive flares. The dashed, solid, and dotted vertical lines denote the flare onset, peak, and decay times respectively.}%\label{fig:?}
\end{figure}
\pagebreak
\subsection{Magnetic-Field Evolution for the Eruptive Flare}

\begin{figure} [h]
\centerline{\includegraphics[trim={.7cm .6cm 1cm 1cm},clip,width=12cm]{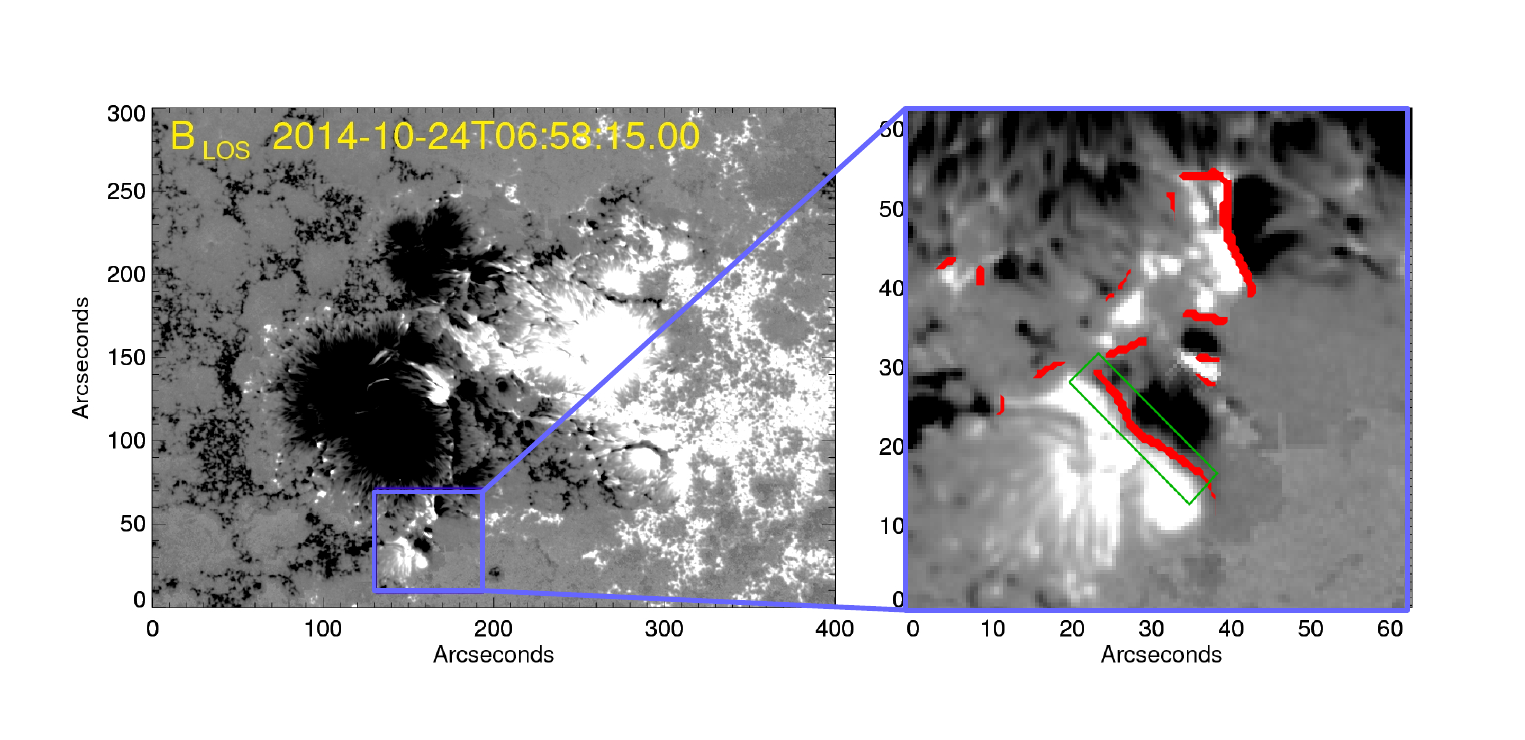}}
\caption{\textit{Left panel}: HMI line-of-sight magnetic field during the eruptive M4.0-class flare. \textit{Right panel}: the line-of-sight magnetic field enlarged from the selected region shown in the left panel. The thick red lines denotes the polarity inversion line. The green boundary in the right panel denotes the selected region within which all the calculations have been done.}
\end{figure}
The eruptive M4.0-class flare occurred away from the core region of AR 12192. The source region of this flare is depicted in Figure 9, where the blue box in the left panel denotes the flaring location identified from the AIA 1600 \AA $ $ images. In this case, the increment in the horizontal magnetic flux (Figure 10) was about 1$\times$10$^{20}$ Mx. Importantly, the change in horizontal magnetic flux for this eruptive flare was about 30 \% from its initial value in the pre-flare stage, whereas those changes for the confined flares range from about 2 \% to 10 \%. Similar to the non-eruptive flares, no significant changes in the vertical component of the magnetic field (upper-right panel of Figure 10) have been found for this flare within the selected region shown by the green boundary in the right panel of Figure 9.

\begin{figure} [H]
  \centering
  \includegraphics[trim={0.32cm 1.5cm 1.75cm 0},clip,width=.464\textwidth]{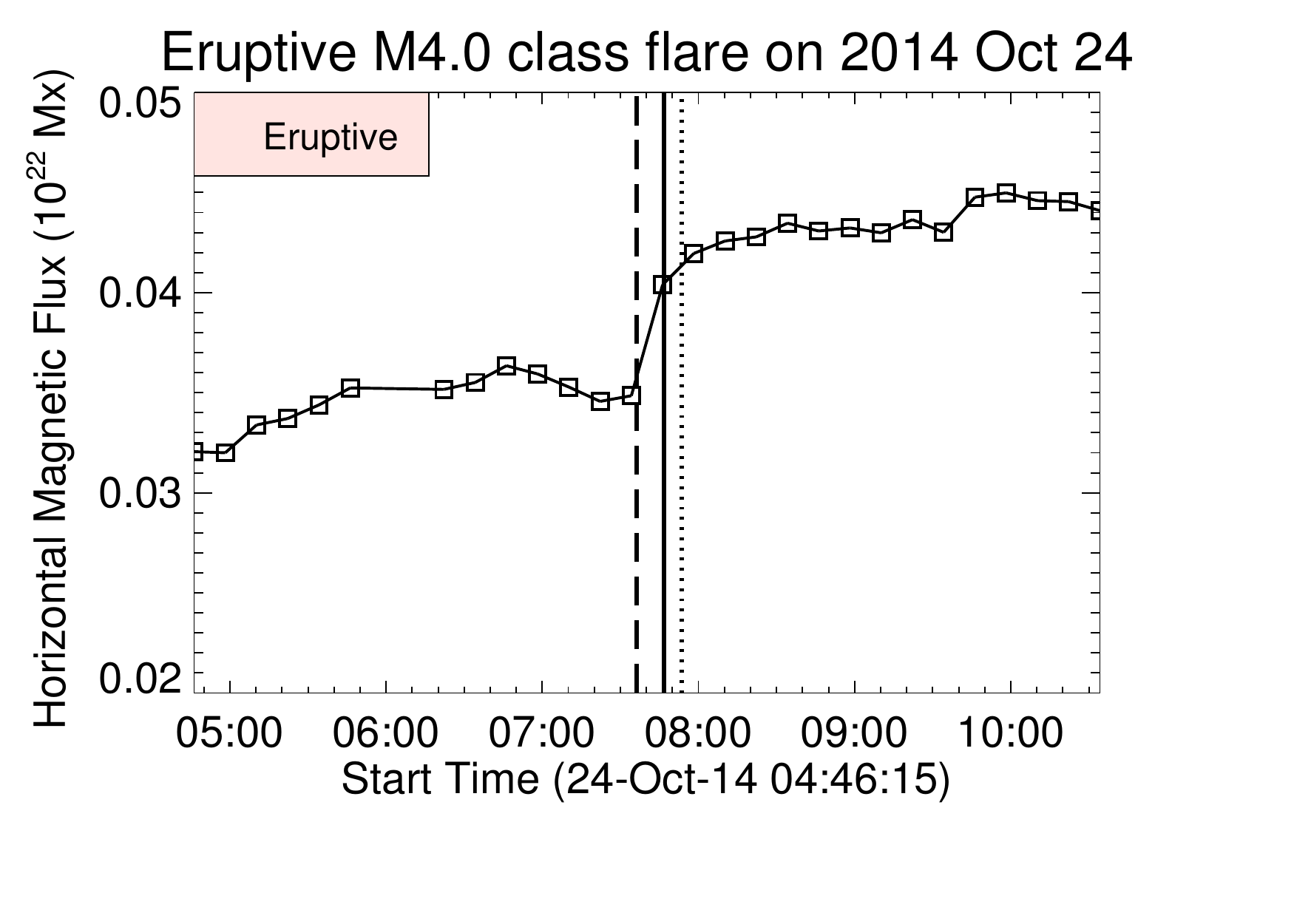}
  \centering
  \includegraphics[trim={0.25cm 1.5cm 0 0},clip,width=.519\textwidth]{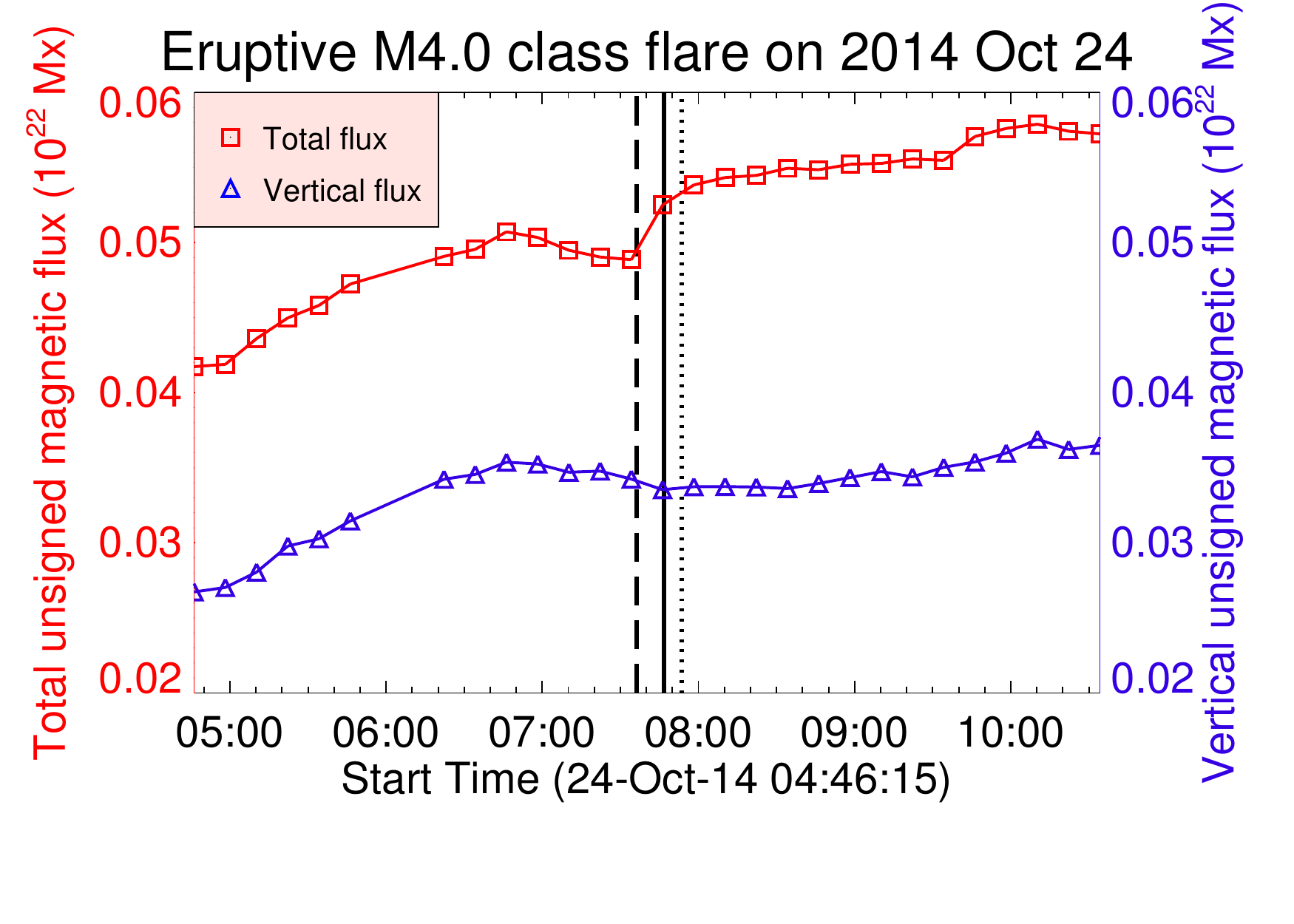}
  
  \centering
  \includegraphics[trim={1.25cm 1.5cm 1.25cm .45cm},clip,width=.458\textwidth]{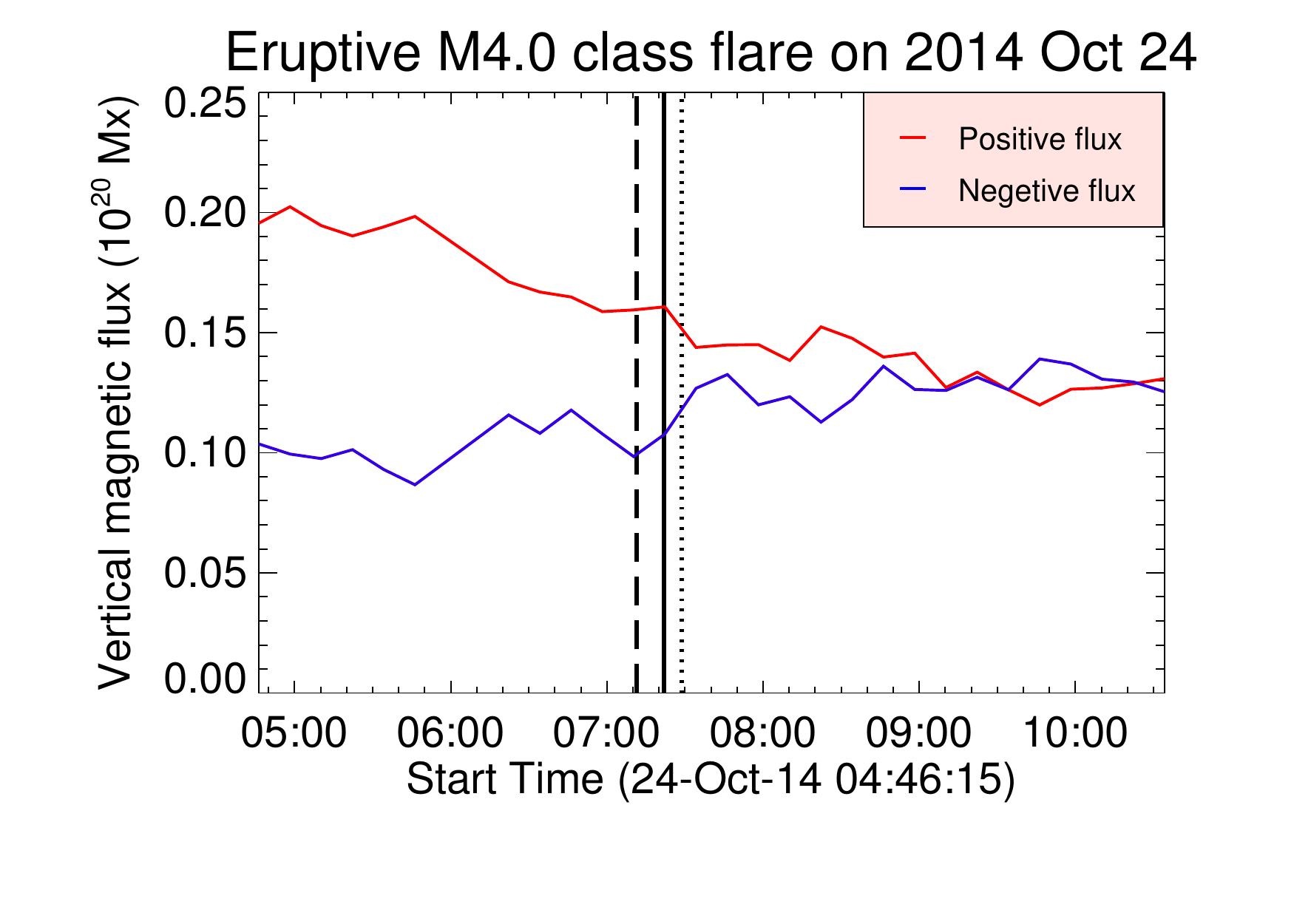}
  \centering
  \includegraphics[trim={0.1cm 1.5cm 0 0},clip,width=.532\textwidth]{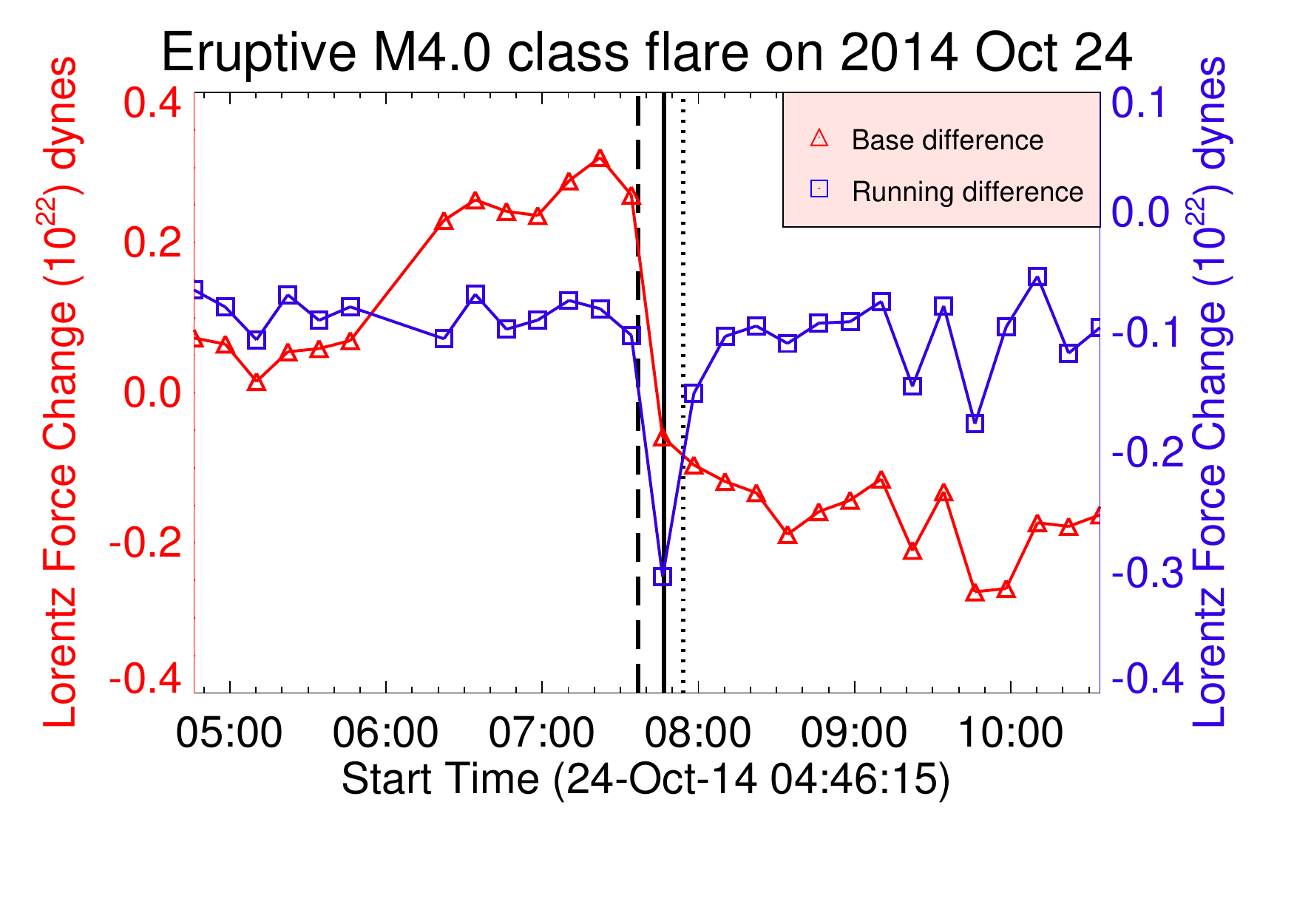}
\caption{Magnetic-field evolution and Lorentz-force changes for the eruptive M4.0-class flare. \textit{Upper-left panel}: the evolution of horizontal magnetic field. \textit{Upper-right panel}: the temporal evolution of the total and vertical magnetic flux. \textit{Lower-left panel}: the temporal evolution of positive and negative magnetic flux. \textit{Lower-right panel}: the changes in radial component of Lorentz force. The dashed, solid and dotted vertical lines denote the flare onset, peak, and decay times respectively. }%\label{fig:?}
\end{figure}
The change in Lorentz force during this eruptive flare was about 0.3$\times$10$^{22}$ dyne (Figure 10). Noticeably, the change in Lorentz force per unit area for this case was about 4040 dyne cm$^{-2}$,  which is almost three times larger than the maximum change found in the four non-eruptive cases. Also, the change in mean horizontal magnetic field during the eruptive flare has been found to be about 135 Gauss, whereas in the confined flares it ranges from about 15 to 35 Gauss only. 
\subsection{Morphological Evolution of AR 12192}
\begin{figure} [!b]
\centerline{\includegraphics[width=12cm]{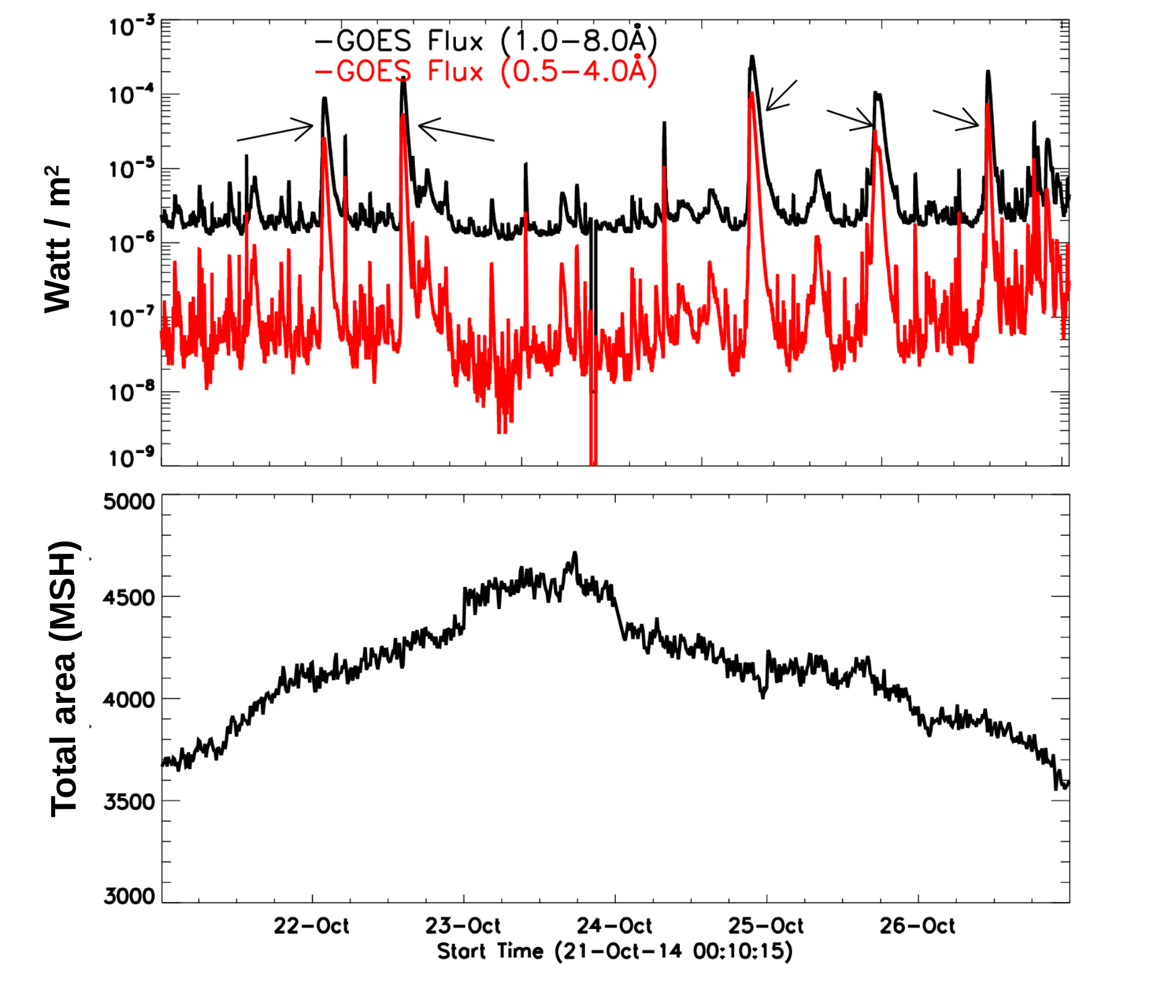}}
\caption{\textit{Top panel}: the temporal profile of GOES X-ray flux during 21 - 26 October 2014. The black arrows in the panel mark the X-class flares. \textit{Bottom panel}: the total area variation of AR 12192 during the same period.}%\label{fig:?}
\end{figure}
AR 12192 underwent significant morphological changes during its disc passage from 17 - 30 October 2014. In between 23 and 24 October it attained the massive size of about 4700 MSH (Millionths of Solar Hemisphere) (Figure 11) and became the largest active region of Solar Cycle 24. From 24 October, both the umbral and penumbral area (Figure 3 in \citealp{Sarkar}) of AR 12192 decayed gradually from its maximum value. The morphological changes of AR 12192 due to this gradual decay of penumbral area are shown in Figure 2, as observed in the high-resolution (0.2 arcsecond) images obtained by MAST. MAST G-band images for four different days (Figure 2) reveal the significant penumbral decay near the core region (marked by black arrows in Figure 2) of AR 12192 where all of the high energetic X-class flares took place.\\

Besides the gradual change, we have also found flare-related abrupt penumbral area decay away from the PIL of the source region of eruptive M4.0-class flare on 24 October 2014. The region of rapid penumbral area decay is illustrated in Figure 12. The blue box in the upper panels shows the bounded region within which the maximum change in penumbral area is observed. The temporal profile of integrated normalized intensity (panel [e] of Figure 12) calculated within the bounded box exhibits a permanent increment after the flare, suggesting the permanent disappearance of the penumbral area after the flare. 
Importantly, this rapid decay in penumbral area was associated with the permanent decay in horizontal magnetic field within the same region. The mean horizontal magnetic field, calculated within the bounded box shown in Figure 12, decreased permanently by almost half of its initial value from about 600 Gauss to 300 Gauss (panel [e] of Figure 12) within less than half an hour during this eruptive flare. This scenario is consistent with the earlier finding of flare-related rapid penumbral decay away from the PIL and the associated permanent decrease in horizontal magnetic field \citep{Xu}. To further quantify this flare-related penumbral area decay, the time evolution of the penumbral area variation is illustrated in Figure 13. The penumbral area is calculated for the contoured sunspot (upper-panel of Figure 13), which was the source region of the eruptive flare. The white and blue contours in the figure mark the umbra--penumbra and penumbra--quiet-Sun boundaries, which are determined by using the normalized intensity thresholds calculated from the cumulative intensity histogram (Figure 4). Noticeably, the temporal profile of penumbral area (lower panel of Figure 13) exhibits a permanent decrement from about 140 to 80 Mm$^2$ within less than half an hour during the flare.

\begin{figure} [H]
\centerline{\includegraphics[width=12cm]{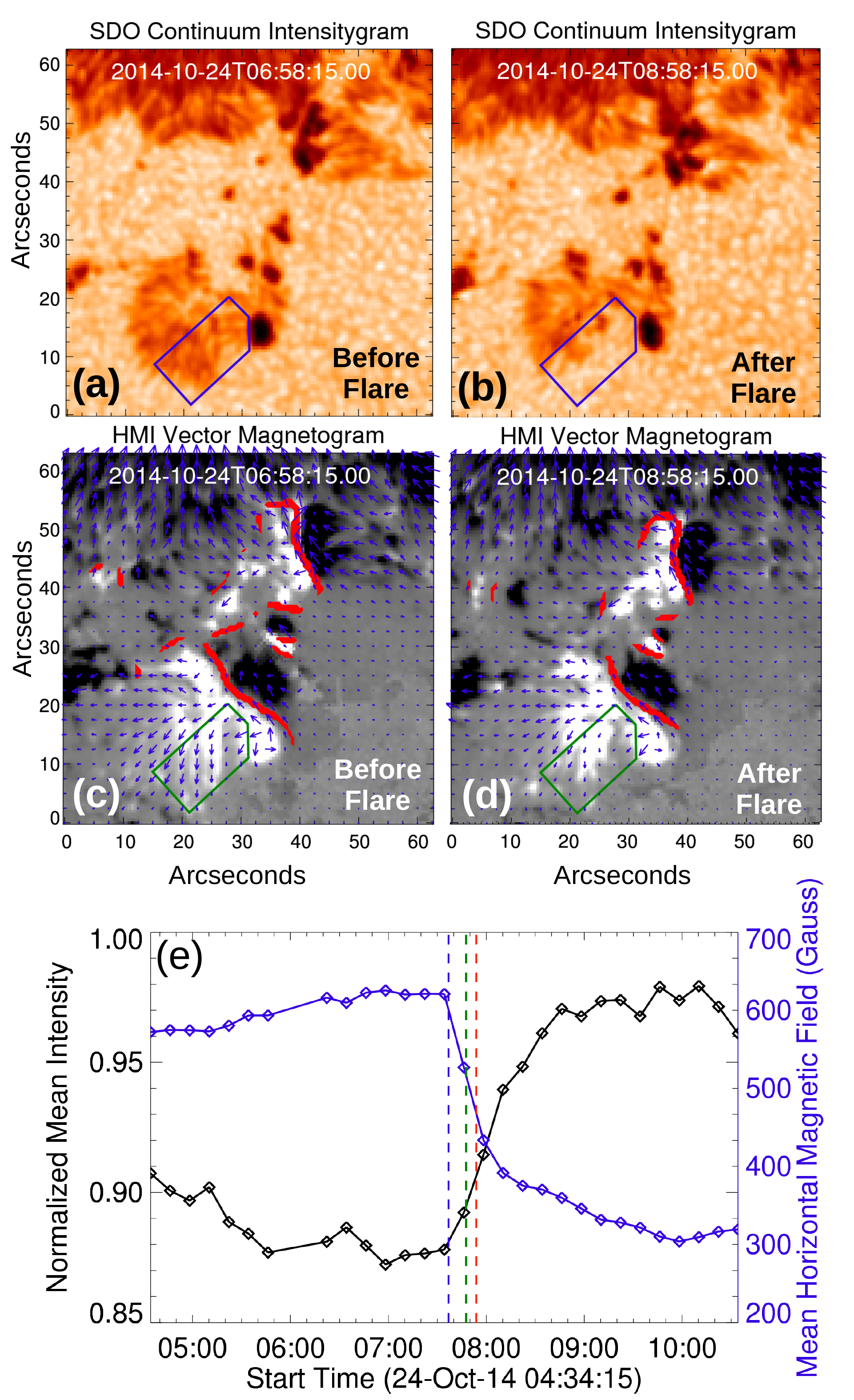}}
\caption{The morphological structure of the penumbral region away from the core region of AR 12192 before (a) and after (b) the eruptive M4.0-class flare respectively. The blue boxes in panels [a] and [b] denote the bounded region where the maximum changes in the penumbral area occurred. Panels [c] and [d] depict the vector magnetograms of the same region shown in panels [a] and [b]. The green boxes in panels [c] and [d] denote the same bounded region marked by the blue boxes in panels [a] and [b]. In the panel [e] the black-solid line represents the temporal variation of normalized intensity and the blue-solid line represents the variation of transverse magnetic field calculated within the box shown in the above panels.}%\label{fig:?}
\end{figure}
\begin{figure}[H] 
\centerline{\includegraphics[width=12cm]{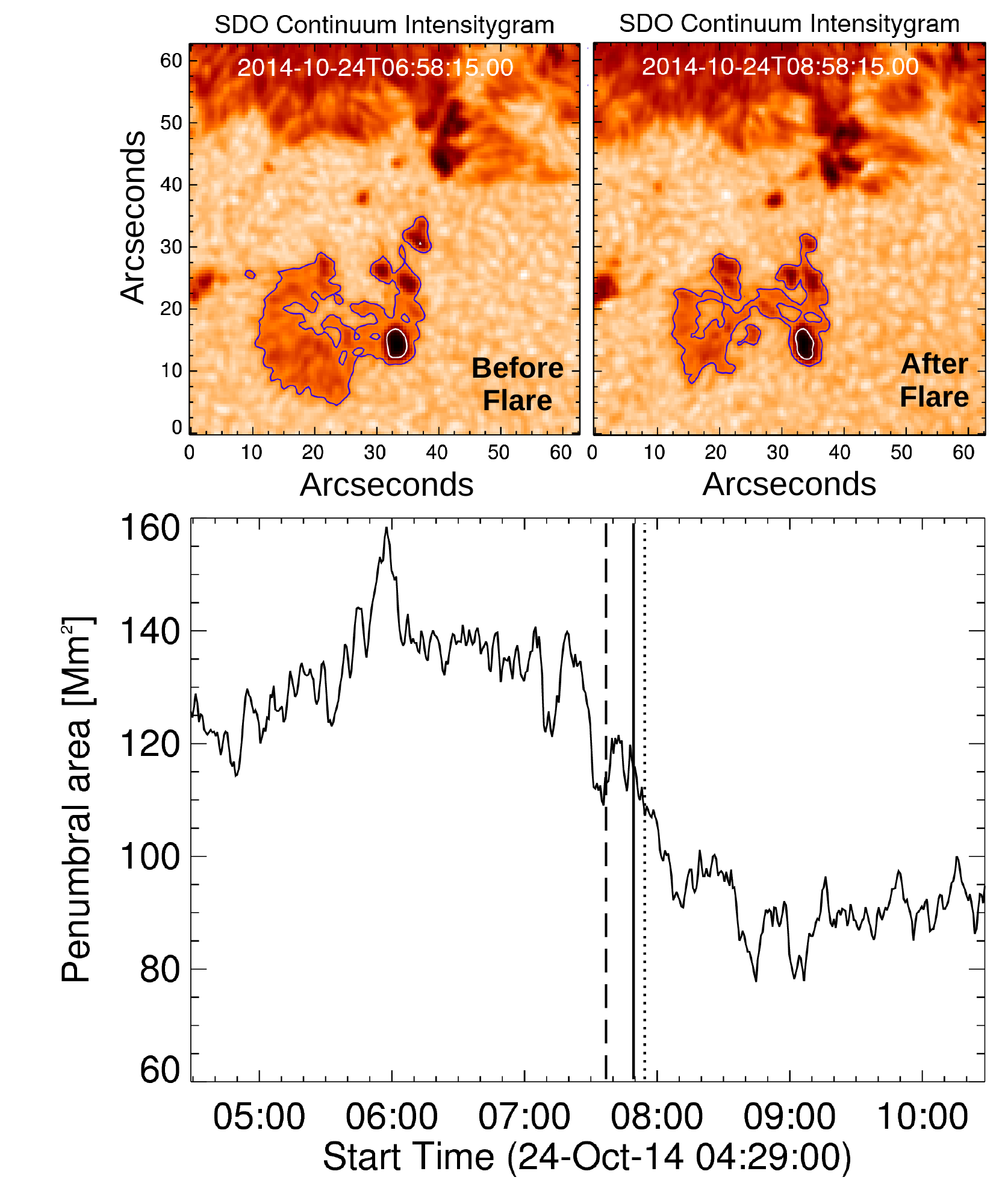}}
\caption{\textit{Lower-panel}: the areal variation of the contoured penumbral region shown in \textit{upper-left} and \textit{upper-right} panels. The dashed, solid, and dotted vertical lines in the \textit{lower-panel} denote the flare onset, peak, and decay times respectively.}%\label{fig:?}
\end{figure}

\subsection{Comparison of Overlying Magnetic-Field Strength for Both the Eruptive and Non-Eruptive Region}

All of the non-eruptive high energetic flares occurred in the core region of AR 12192 (Figure 3). However, the eruptive flare occurred away from the core region (Figure 14 a,b). From the study of extrapolated magnetic field, we have found that the overlying magnetic-field strengths were different for the eruptive and non-eruptive flaring regions. 
\begin{figure}[H]
\centerline{\includegraphics[trim={0 0 1cm .5cm},clip,width=12cm]{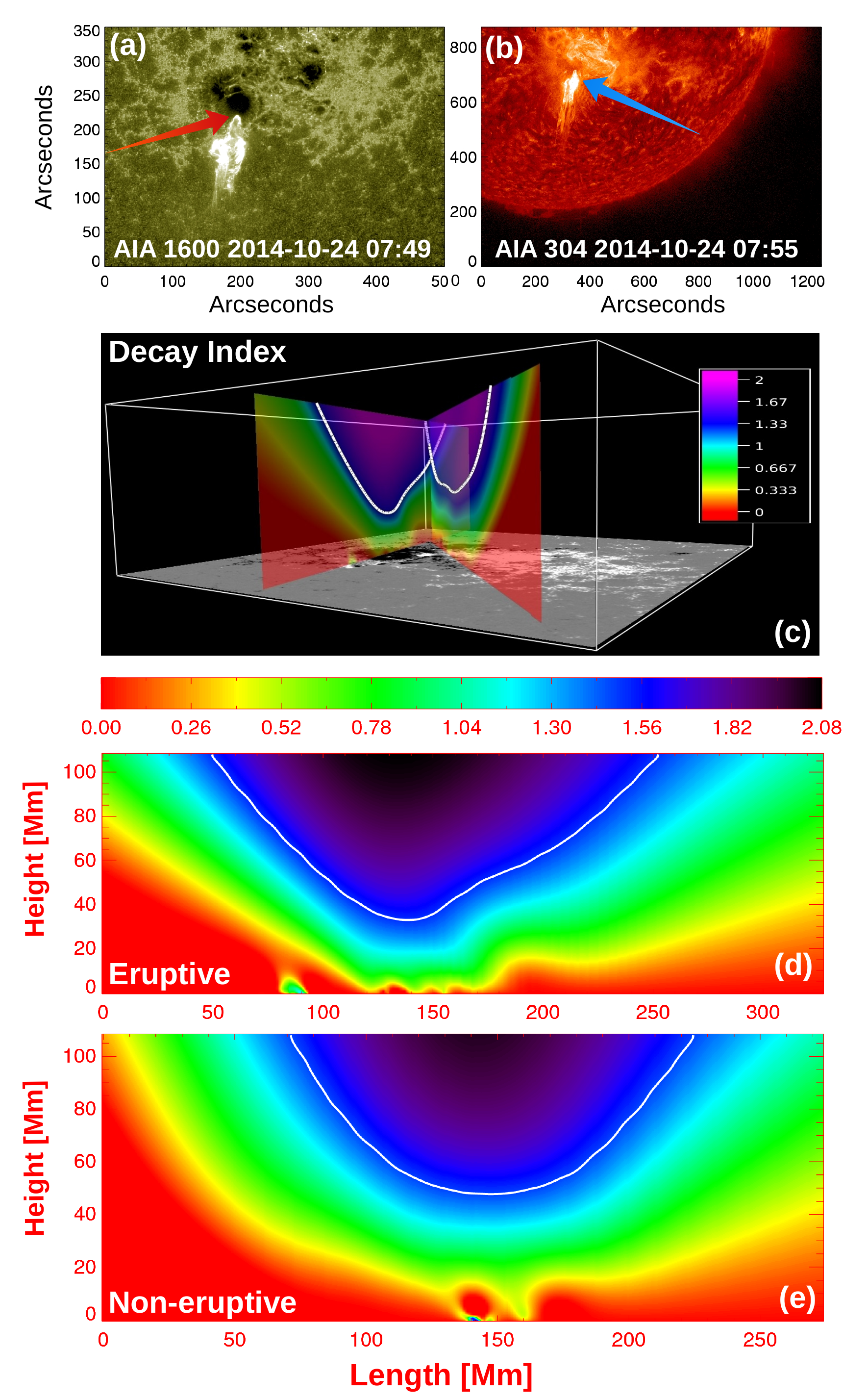}}
\caption{The red and blue arrows in panels [a] and [b] mark the source location of the eruption on 24 October 2014 in AIA 1600 \AA\ and 304 \AA\ images respectively. Panel [c] depicts the two decay-index profiles drawn over the HMI line-of-sight magnetogram. The right-hand decay index profile in panel [c] is along the main PIL of the non-eruptive core region and the left-hand decay index profile is drawn over the eruptive flaring region. The two decay index profiles shown in panel [c] are illustrated separately in panels [d] and [e] respectively.}%\label{fig:?}
\end{figure}
In order to quantify the overlying magnetic-field strength, we have calculated the decay index over the whole AR. To compare the decay-index profile for both the eruptive and non-eruptive region of AR 12192, two 2D profiles of decay index are plotted in Figure 14, where one of them is along the main PIL of the core region, which was the non-eruptive zone, and the other one is over the source region of the eruptive flare, which occurred away from the core region. Interestingly, the two decay index profiles show different depths in the contour drawn for 1.5 value, which is believed to be the critical decay index for the onset of the torus instability. The critical decay index was achieved at a height about 52 Mm over the non-eruptive core region of AR 12192, whereas the decay index decreased to the critical value (1.5) at height 35 Mm over the eruptive region. This means that the external magnetic field decays faster over the erupting region than the non-eruptive one. This result further suggests that the strength of the overlying magnetic field plays a decisive role in determining whether the AR will be CME-productive or not.\\

\section{Summary and Conclusion}
Table 2 summarizes our results based on a comparative study of the eruptive and non-eruptive flares produced by the largest AR of Solar Cycle 24 : AR 12192.     
\begin{table}[!h]
%\caption {Comparison between the eruptive and non-eruptive flares produced by AR 12192 } \label{tab:title}
\begin{adjustbox}{max width=\textwidth}
\label{tab2}

\caption {Comparison between the eruptive and non-eruptive flares produced by AR 12192 } \label{tab:title1}
\begin{tabular}{lccccc} % define the column alignment
                                  % l: left, c: center, r: right
                                  % @{.} replace the inter-column by a .
  \hline
  \hline

%Date               & 2014/10/22   & 2014/10/22   & 2014/10/24 & 2014/10/24   & 2014/10/25 \\
Date               & 22 October & 22 October & 24 October & 24 October & 25 October \\
\hline
GOES peak time [UT]          &   01:59    &    14:06      &  07:48      &   21:15       &  17:08 \\
\hline
GOES flare class        & M8.7         & X1.6         & M4.0       & X3.1         & X1.0\\
\hline
Nature of eruption        & Non  & Non &    & Non  & Non \\
                   & eruptive & eruptive &  Eruptive  & eruptive & eruptive\\ 
\hline
Change in horizontal\\
magnetic flux [Mx] & $\approx$ 3$\times$10$^{20}$ & $\approx$ 5$\times$10$^{20}$ & $\approx$  1$\times$10$^{20}$  &$\approx$  5$\times$10$^{20}$ & $\approx$ 7$\times$10$^{20}$\\

\hline
Percentage of change in\\
horizontal magnetic flux& $\approx$ 2\% & $\approx$ 2.5\% & $\approx$ 30\%&$\approx$ 5\% & $\approx$ 10\%\\

\hline
Change in mean \\
$B_{\rm h}$ [Gauss] & $\approx$ 15 & $\approx$ 15 & $\approx$ 135  & $\approx$ 25 &$\approx$ 30\\
     
\hline
Change in Lorentz \\
force [dyne] & $\approx$ 2.6$\times$10$^{22}$  & $\approx$ 1.5$\times$10$^{22}$ & $\approx$ 0.3$\times$10$^{22}$  & $\approx$ 1.7$\times$10$^{22}$  & $\approx$ 2$\times$10$^{22}$ \\

\hline
Change in Lorentz force  \\
per unit area [dyne cm$^{-2}$] & $\approx$ 1390  &$\approx$ 390 &$\approx$ 4040 & $\approx$ 890 &$\approx$ 960\\
%\hline
%\textbf{Photospheric flux} & \textbf{Not}&\textbf{Not} &    &\textbf{Not}& \textbf{Not}\\ 
%\textbf{cancellation} &\textbf{significant} &\textbf{significant} &\textbf{Significant} & \textbf{significant}&\textbf{significant}  \\

\hline
Morphological & Not & Not &    &Not & Not \\ 
change& significant&significant &Significant  &significant & significant\\
\hline
Overlying magnetic-\\
field strength& Strong & Strong &  Weak  & Strong & Strong\\ 

 \hline
  \hline

\end{tabular}
\end{adjustbox}
\end{table}

By comparing the magnetic characteristics for both the eruptive and non-eruptive flares, we conclude that the eruptive flare left a significant magnetic imprint on the solar photosphere, whereas the photospheric magnetic field changes were comparatively small in the case of the confined flares. In contrast to the eruptive flare, the confined flares exhibited very weak changes in horizontal magnetic-field and Lorentz force per unit area (Table 2). This scenario is consistent with the flare-related momentum balance condition where the Lorentz-force impulse is believed to be proportional to the associated CME momentum \citep{Fisher,ShouWang}. Thus, if a flare is not associated with a CME then the Lorentz-force impulse during that confined flare is expected to be smaller. Our results from the analysis of selected four non-eruptive flares are in agreement with this expectation. All of the five flares of our dataset showed abrupt and permanent changes in photospheric magnetic field, which is a common feature in large flares \citep{Wang2012,Sun2012}. However, the results presented in the current article suggest the changes to be larger in magnitude for the eruptive flares in comparison to the confined ones. Therefore, the flare-related temporal changes in magnitude of the transverse magnetic field and Lorentz-force impulse appeared to be well correlated with the nature of eruptions. The weak changes in the magnetic characteristics during the four non-eruptive flares support the earlier results of \citet{Sun2015} where they compared the photospheric changes during one non-eruptive X3.1-class flare produced by AR 12192 with those during two eruptive X5.4 and X2.2-class flares produced by different ARs, i.e. AR 11429 and AR 11158 respectively. However, we have followed a different approach where we compared the photospheric magnetic environments of the four highly energetic non-eruptive flares with those for the eruptive flare originated from the same AR (AR 12192). Therefore, our analysis is free from the bias of other factors, such as the scale size and complexity of the different ARs, which are believed to play important role in the eruptions.

The significant growth in the AR area is reflected in the magnetic energy content of AR 12192, as it produced six highly energetic X-class flares within an interval of one to two days during its disc passage. This implies that the huge size of AR 12192 may be responsible for its energy storage being sufficient to trigger the recurrent highly energetic flares. The rapid penumbral-area decay observed during the eruptive M4.0-class flare reveals how the strong Lorentz force impulse can shape the dynamics of the solar eruptions and cause morphological changes in the photospheric features. The physical explanation behind these flare-related photospheric morphological changes is given in \citet{Xu}. They showed that during the flare, due to the magnetic-pressure imbalance between the reconnection site and the outer atmosphere, the Lorentz force away from the PIL acts in the upward direction to lift up the magnetic field lines from the outer higher magnetic-pressure region towards the reconnection site. Therefore, away from the PIL the more horizontal field lines in the preflare stage become more vertical after the flare and release the magnetic pressure due to the upward impulse of the Lorentz force, resulting in a decrement in horizontal magnetic-field strength and disappearance of penumbral area. However, in the case of the non-eruptive flares we did not find any noticeable flare-related permanent and abrupt changes in the umbral or penumbral area. The presence of weak Lorentz-force impulse and different restructuring in the magnetic-field topology in the confined flares may be attributed to the insignificant morphological changes found in those cases.   

Comparison of the overlying coronal magnetic field environment in the pre-flare stages reveals that the gradient of overlying magnetic-field strength decayed faster over the eruptive region of AR 12192 as compared to that over the non-eruptive zone, suggesting the overlying field strength was stronger over the non-eruptive core area of the AR than the eruptive zone. This result is similar to the earlier finding by \citet{cheng} where they did a comparative study of the critical decay index (1.5) height for the onset of the torus instability over both the eruptive and non-eruptive regions belonging to same AR (NOAA AR 10720). Although our study suggests the importance of the background magnetic field strength for the confined behavior of the non-eruptive flares, a recent study \citep{Zhang2017} reveals that the complex flare loops associated with the non-eruptive flares originating from AR 12192 may also be responsible for the confinement.    

In short, our comparative study of both the eruptive and non-eruptive flares produced from AR 12192 suggests that, although the flare-related permanent and abrupt changes in photospheric magnetic field and Lorentz forces are a common feature in large flares, the magnitude of those changes is smaller in the case of the confined flares compared to the eruptive ones. We conclude that the highly energetic flares leave a magnetic imprint on the solar photosphere that carries information on the nature of eruption. In addition, the comparative study of overlying coronal magnetic field environments for both the confined and eruptive flares reveal that the decay rate of the overlying magnetic-field strength can be used as a key parameter to determine the CME productivity of the AR. Nevertheless, our analysis is limited to a small number of events. In the future, we plan to extend our work to several confined and eruptive events using the newly available high cadence (135 seconds) HMI vector magnetic-field data.

%%%%%%%%%%%%%%%%%%%%%%%%%%%%%%%%%%%%%%%%%%%%%%%%%%%%%%%%%%%%%%%%%%%%%%%%%%%
%% Appendix
%
% \appendix   

%%%%%%%%%%%%%%%%%%%%%%%%%%%%%%%%%%%%%%%%%%%%%%%%%%%%%%%%%%%%%%%%%%%%%%%%%%%
%% Acknowledgements
%
\begin{acks}
Ranadeep Sarkar would like to thank Sajal Kumar Dhara for useful discussion on this work. We acknowledge NASA/SDO and the AIA and HMI science teams for data support. We also acknowledge  the contribution of the MAST team members for the acquisition of the data. The use of the visualization software \textsf{VAPOR (www.vapor.ucar.edu)} for generating relevant graphics is sincerely acknowledged. This work was (partly) carried out by using Hinode Flare Catalogue \textsf{(hinode.isee.nagoya-u.ac.jp/flare$\_$catalogue/)}, which is maintained by ISAS/JAXA and Institute for Space-Earth Environmental Research (ISEE), Nagoya University.  
 \end{acks}

 \section*{Disclosure of Potential Conflicts of Interest}
 The authors declare that they have no conflicts of interest.

\pagebreak

%%% %%%%%%%%%%%%%%%%%%%%%%%%%%%%%%%%%%%%%%%%%%%%%%%%%%%%%%%%%%%
%% Bibliography
%
% Using BibTeX
%Random citation \cite{2015} embeddeed in text.

% \bibliographystyle{spr-mp-sola}
% \bibliography{references}  
%
% Without BibTeX 
% \begin{thebibliography}{}
% \bibitem[\protect\citeauthoryear{Author}{Year}]{key}
%   <bibliographical entry>
%
% \bibitem[\protect\citeauthoryear{}{}]{}
%   
%  
% \end{thebibliography}

\end{article} 
\end{document}